\documentstyle{article}

\def\afour{
\setlength{\topmargin}{0mm}
\setlength{\headheight}{0mm}
\setlength{\headsep}{0mm}
\setlength{\textwidth}{6in}
\setlength{\textheight}{248mm}
\setlength{\oddsidemargin}{.25in}
\setlength{\evensidemargin}{.25in}
}

\newcommand\eq[1]{Eq.~(\ref{#1})}
\newcommand\eqs[2]{Eqs.~(\ref{#1}) and (\ref{#2})}
\newcommand\eqss[3]{Eqs.~(\ref{#1}), (\ref{#2}) and (\ref{#3})}
\newcommand\eqsss[4]{Eqs.~(\ref{#1}), (\ref{#2}), (\ref{#3})
and (\ref{#4})}
\newcommand\eqssss[5]{Eqs.~(\ref{#1}), (\ref{#2}), (\ref{#3}),
(\ref{#4}) and (\ref{#5})}
\newcommand\eqst[2]{Eqs.~(\ref{#1})--(\ref{#2})}

\newcommand\rfrac[2]{\left(\frac{#1}{#2}\right)}

\newcommand{\sub}[1]{_{\mbox{\scriptsize#1}}}
\newcommand{\su}[1]{^{\mbox{\scriptsize#1}}}

\newcommand\ee{\end{equation}}
\newcommand\be{\begin{equation}}
\newcommand\eea{\end{eqnarray}}
\newcommand\bea{\begin{eqnarray}}

\newcommand\yr{\,\mbox{yr}}
\newcommand\sunit{\,\mbox{sec}}
\newcommand\munit{\,\mbox{m}}

\newcommand\cm{\,\mbox{cm}}
\newcommand\km{\,\mbox{km}}
\newcommand\TeV{\,\mbox{TeV}}
\newcommand\GeV{\,\mbox{GeV}}
\newcommand\MeV{\,\mbox{MeV}}
\newcommand\keV{\,\mbox{keV}}
\newcommand\eV{\,\mbox{eV}}
\newcommand\Mpc{\,\mbox{Mpc}}

\newcommand\mone{^{-1}}
\newcommand\mtwo{^{-2}}
\newcommand\mthree{^{-3}}
\newcommand\mfour{^{-4}}
\newcommand\mhalf{^{-1/2}}
\newcommand\mthreehalf{^{-3/2}}
\newcommand\mthird{^{-1/3}}

\newcommand\half{^{1/2}}
\newcommand\threehalf{^{3/2}}
\newcommand\third{^{1/3}}
\newcommand\twothird{^{2/3}}
\newcommand\quarter{^{1/4}}

\newcommand\msun{M_\odot}
\newcommand\mpl{m_{Pl}}

\newcommand\del{{\mbox{\boldmath $\nabla$}}}
\newcommand\bfk{\mbox{\bf k}}
\newcommand\bfr{\mbox{\bf r}}
\newcommand\bfv{\mbox{\bf v}}
\newcommand\bfw{\mbox{\bf w}}
\newcommand\bfe{\mbox{\bf e}}
\newcommand\bfx{\mbox{\bf x}}

\newcommand\bfu{\mbox{\bf u}}
\newcommand\sk{_{\mbox{\scriptsize \bf k}}}

\newcommand\pa{\partial}
\newcommand\pdif[2]{\frac{\pa #1}{\pa #2}}

\newcommand\lsim{\mathrel{\rlap{\lower4pt\hbox{\hskip1pt$\sim$}}
    \raise1pt\hbox{$<$}}}
\newcommand\gsim{\mathrel{\rlap{\lower4pt\hbox{\hskip1pt$\sim$}}
    \raise1pt\hbox{$>$}}}

\newcommand\diff{\mbox d}

\def\e{\mbox e}

\def\n{\mbox n}

\def\p{\mbox p}

\def\A{\mbox A}
\def\B{\mbox B}
\def\C{\mbox C}
\def\D{\mbox D}

\def\H{\mbox H}

\def\calp{{\cal P}}
\def\calr{{\cal R}}

\def\ebar{\bar{\e}}

\def\Hethree{{}^3{\mbox {He}}}
\def\Hefour{{}^4{\mbox {He}}}
\def\Liseven{{}^7{\mbox {Li}}}

\afour

\begin{document}

\begin{titlepage}
\begin{flushright}
LANC-TH 9319\\
astro-ph 9312022\\
November 1993\\
\end{flushright}\
\vspace{.1cm}
\begin{center}
{\huge \bf INTRODUCTION\\
\vspace{.35cm}
TO \\
\vspace{.6cm}
COSMOLOGY}\\
\vspace{.8cm}
{\large Lectures given at the\\
\vspace{.2cm}
Summer School in High Energy Physics and Cosmology\\
\vspace{.3cm}
ICTP (Trieste) 1993}\\
\vspace{1cm}
{\large David H.~Lyth\\}
\vspace{.4cm}
{\em School of Physics and Materials,\\ Lancaster University,\\
Lancaster LA1 4YB.~~~U.~K.}\\
\end{center}
\vspace{1cm}
\begin{abstract}
\baselineskip=24pt
\noindent
These notes form an introduction to cosmology with special emphasis on
large scale structure, the cmb anisotropy and inflation.
In some places a basic familiarity with particle physics is assumed, but
otherwise no special knowledge is needed.
Most of the material in the first two sections
can be found in several texts, except that the
discussion of dark matter and the cosmological constant is more up to
date. Most of that
in the remaining sections can be found in a review of structure
formation and inflation done with
Andrew Liddle, which describes original work by various authors
including ourselves and Ewan Stewart. The reader is referred to these
works for more detail, and a very complete list of references.
\end{abstract}

\end{titlepage}

\newpage
\tableofcontents
\newpage

\section{The Recent Universe}

These notes form an introduction to cosmology with special emphasis on
large scale structure, the cmb anisotropy and inflation.
In some places a basic familiarity with particle physics is assumed, but
otherwise no special knowledge is needed.
Most of the material in the first two sections
can be found in several texts \cite{kotu}, except that the
discussion of dark matter and the cosmological constant is more up to
date. Most of that
in the remaining sections can be found in a review of structure
formation and inflation done with
Andrew Liddle \cite{LL2}, which describes original work by various authors
including ourselves and Ewan Stewart. The reader is referred to these
works for more detail, and a very complete list of references.

According to current thinking, the history of the observable universe
broadly divides into three stages. First there is an {\it inflationary era},
when the energy density is dominated by the potential of a scalar field.
Then there is a {\it radiation dominated era} when the energy density
is dominated
by relativistic particles, which are called `radiation' by cosmologists.
Finally, lasting till the present epoch,
there is a {\it matter dominated era} when the energy density is
dominated by the mass of non-relativistic particles, which are called
`matter'.
This first lecture concerns the latter part of the matter dominated
era, and in particular the present epoch.

Unless the contrary is implied by the specified units,
I set $\hbar=c=k_B=1$.

\subsection{How one decides what is true}

Before beginning let us address an important issue.
More than most scientists, cosmologists tend to be asked `how do you
know?' The answer is that one proceeds in the same as in other areas of
physics and related
sciences. One starts with a {\it model}, which is
a statement about the nature of physical reality, plus a theory
formulated in terms of equations. Then one calculates (`predicts')
the outcome of observations and experiments that have been performed, or
better of ones that are about to be performed. If the agreement is
impressive enough everyone agrees that the model is correct.

 This sequence {\it cannot} be
reversed, that is one cannot rigorously work back to deduce the
model from observation. To put it the
other way round, a model can strictly speaking only be falsified, not verified.
Logically, two completely different models could
both agree with observation but that doesn't seem to happen, though of
course an old model  can be seen to be a limiting cases of a better one.

So the answer to the question `how do you know' begins with the
admission that a guess has been made. At any one time, though,
parts of the picture are universally accepted as
correct, because they agree with such a huge body of observation.
Other parts of the picture are just coming into focus, and agreement
about the truth lies in the future. The scope of the agreed part
is dictated not by logic, but by consensus.

Except for the mathematical theory, all
of this is not really special to science, but is rather just the way
that human beings go about things. For instance, at any given epoch
during the centuries that the
geography of the earth's surface was being
established in the West, parts of it were universally accepted because
the accounts of many people agreed (or sometimes because just one of two
surveyors had done what looked like a good job). At the same time,
parts of it were still in dispute. The portion of the earth's surface
about which there was agreement was dictated not by logic, but by
consensus.

Because our knowledge has been arrived at by comparing a large
body of ideas with a large body of observation, it is not usually
possible to point to a particular observation as the reason for
believing a particular fact. For this reason, as well as because
of the introductory nature of the lectures, I will often state
a fact without giving specific justification. On the
other hand, by the time we have finished it will hopefully be clear that
the picture I put forward leads to agreement with an impressive
body of observation, and that this picture could hardly be subjected to
drastic alteration without spoiling this agreement.

\subsection{The geography of the observable universe}

Astronomers use the unit 1\,pc$=3.26$\,light years$=3.09\times 10^{16}
$\,metres. For cosmology the megaparsec,
$1\Mpc=10^6$\,pc is appropriate. The unit of mass is the
solar mass $\msun=1.99\times10^{33}$\,g.

Stars have mass in the range roughly $1$ to $10\msun$.
They are found only in
{\it galaxies} may be regarded as the
basic building blocks of the universe, with masses ranging from
maybe $10^6\msun$ (dwarf galaxies) to $10^{12}\msun$ (large galaxies
like our own). A galaxy typically has a luminous centre containing nearly all
of the stars, and a dark halo of unknown composition which extends of
order 10 times as far and contains of order 10 times as much mass.
In many (all?) cases there is a nucleus
consisting of a black hole which is gobbling up stars and gas
and emitting radiation. If the nucleus is the dominant feature
it is called an AGN (active galactic nucleus),
 examples being Seyfert galaxies and
quasars. (Astronomers tend to
using
different names for different examples of the same type object.
The reason, of course,
is that they don't know at first that they
{\it are} examples of the same type of object.)

In round figures, large galaxies
like our own have a size of .1\,Mpc (including the dark halo)
and are of order 1\,Mpc apart. Many galaxies belong to
gravitationally bound clusters containing from two to $\sim 1000$
galaxies. Small clusters are usually called groups. Big
clusters, of order 10\,Mpc in size, are the biggest gravitationally bound
objects in the universe. There do exists, though, `superclusters' with size
of order 100\,Mpc, which are conglomerations of galaxies and clusters
representing regions of space which have higher than average density.
Presumably they will
become gravitationally bound at some time in the future. On the scale of
$100\Mpc$ there also seem to be sheetlike and filamentary structures, as
well as voids containing hardly any galaxies.

On scales bigger than $100\Mpc$ the distribution of matter in the
universe is known to be very homogeneous, both from direct observation of
the galaxies and from the isotropy of the microwave background.
To be precise, if one throws down at random a sphere with radius
$R$ and measures its mass $M$, then the {\it rms} variation
$\Delta M/M$ is a decreasing function of $R$, which is of order
1 at $R=10\Mpc$ and of order .1
at $R=100\Mpc$

The biggest distance we can observe is of order $10^4\Mpc$,
the distance that light has travelled since the Big Bang.
The sphere around us with this radius is thus the {\it observable
universe}. As far as we can tell, distant
parts of the observable universe are much
like our own part, with the same kind of
galaxies, clusters and so on.
{}From the fact that
the microwave background anisotropy $\Delta T/T$ is of order
$10^{-5}$, one can deduce that
$\Delta M/M\lsim10^{-5}$ on scales $R$ comparable with
the size of the observable universe.

\subsubsection*{What's beyond the observable universe?}

What is the universe like outside the patch we observe?
Since the universe
is very homogeneous and isotropic in the observed patch,
it is reasonable to suppose that it
remains so for at least a few orders of magnitude further.
But what about happens after that?

Until the last decade or so, the prevailing assumption seems to have
been that the entire universe is homogeneous and isotropic, like the
bit that we observe. To avoid the
embarrassing question of what would then be beyond the edge of
the `entire' universe, this view requires that one believes in
a `closed' universe, which is allowed in the context of
general relativity. Such a universe is the three dimensional analogue of
a sphere, and as I discuss later on this lecture its spatial
`curvature' could be detectable observationally.

More recently,
the favoured view has been that as one examines larger
and larger scales, the universe becomes steadily more inhomogeneous
and anisotropic \cite{CHAOTIC}.
 Going the other way, if one takes an overall view of a
patch of the universe many orders of magnitude bigger than our own,
it has a non-fractal
nature; as one goes down to smaller and smaller scales, it looks more
and more homogeneities, as opposed to the `fractal' situation where
new inhomogeneities would reveal themselves at every step.
On this view, the observable universe is {\it extremely}
homogeneous because it is a small part of a much large patch which
is {\it roughly} homogeneous and isotropic.
As I shall discuss in Lecture 6, this viewpoint is
normally discussed within the context of inflation which indeed makes
it useful and rather compelling, but one should understand that it
is a separate hypothesis, not related {\it per se} to the inflationary
hypothesis.

On this second view, there can be yet another hierarchy as one moves beyond
the roughly homogeneous and isotropic patch.
One could encounter regions where the universe is radically different,
for instance contracting instead of expanding, or with
different laws of physics, corresponding to different solutions of some
super-theory. One would presumably still like the `entire' universe
to be closed, but on this second view it would be analogous to
the surface of an
extremely deformed `sphere' so that no hint
of the global behaviour would be evident from observations of our own
region.

\subsection{The isotropically expanding universe}

On the large scales where it is homogeneous, the universe is also
expanding isotropically. That is, the distance between any pair of
galaxies separated by more than $100\Mpc$ is proportional to a universal
{\it scale factor} $a(t)$, the same for every pair.
The {\it Hubble parameter} is defined by $H=\dot a/a$, where the dot
is the time derivative.

A subscript $0$ is generally used to denote quantities evaluated
at the present epoch.
It is convenient to set $a_0=1$, so that $a(t)$ is simply the
size of any comoving
region (one moving with the galaxies) relative to its present size.
The present value of $H$, denoted by $H_0$ is called the Hubble
constant. It is traditionally measured by observing the redshift
$z\equiv\Delta \lambda/\lambda$
of galaxies receding from us with velocity $v\ll 1$.
The velocity of such a galaxy is given by $v=Hr$, and its
redshift is just the non-relativistic Doppler shift $z=v$, leading to
Hubble's law
\be
z(=v)=H_0 r_0
\label{1}\ee

Hubble's law is well established because {\it relative}
distances are
easy to establish. All one has to do is find a
`standard candle', that is a type of object
(say as star of a given type) of which all examples have
practically the same luminosity.
Then its apparent luminosity will
vary with (distance)${}\mthree$, and so measure relative
distances.\footnote
{To establish that an object {\it is} a standard candle one should
ideally understand it theoretically, but in practice it is fairly safe
to assume that a sufficiently
distinctive object is standard.
As a check one can compare two different candles, observing pairs of
them which are known for
to be in the same region of space because, for instance, they are in the
same galaxy or star cluster. If they both have luminosity proportional
to (distance)${}\mthree$, either they are both standard or they conspire
to be non-standard in the same way.}
On the other hand
to fix $H_0$ which is the constant of proportionality one has to
know the luminosity of some object, which is much harder to do.
Different estimates give
$H_0$ in the range $40$ to $100\km\sunit\mone\Mpc\mone$, and it is usual
to define a quantity $h$ by
\be
H_0 = 100 h
\km\sunit\mone\Mpc\mone
\label{2}\ee
Thus, the redshift determination gives $.4<h<1$.
As one might expect, $H_0$ enters into many other equations besides
Hubble's law, so the redshift determination is {\it not} the only
method of determining $H_0$. I will suggest a `best fit' value and
error later on.

Since $H\equiv\dot a/a$, the
time taken for the universe to expand by an appreciable amount
is of order the {\it Hubble time}
\be
H_0\mone=9.76h\mone \times10^9\yr
\label{4}\ee
During a Hubble time, light travels a distance of order the
{\it Hubble distance},
\be
H_0\mone= 2998 h\mone \Mpc
\label{3}\ee
Hubble's law applies only to galaxies whose distance is much less
than the Hubble distance, it is based on the non-relativistic Doppler
shift which requires $v\ll 1$.
As we will discuss later, much more distant galaxies are also observed,
the radius of the observable universe being of order
the Hubble distance. Thus we are at the moment discussing only the
{\it nearby} part of the observable universe.
The universe has expanded by a negligible amount since
light from the nearby universe was emitted, but by a significant amount
since light from the distant universe was emitted.

\subsubsection*{The Big Bang}

At the present epoch the Hubble time is of order $10^{10}\yr$.
In what follows we shall extrapolate back to an era when the universe is
very hot and dense, and the Hubble time is only a tiny fraction of a second.
This era is popularly known as the Hot Big Bang, but one should not be
misled by the language. Since the early universe is even more
homogeneous and isotropic than the present one the expansion cannot be
said to originate from a central point. It is certainly not an
explosion, which by definition is driven by a pressure gradient.
Thus the undoubted fact of the Big Bang does not explain the expansion
of the universe; it has to be laid down at the very beginning.

When is the beginning? Presumably it lies at the
{\it Planck epoch}, when the Hubble time is of order the Planck
time
\be
t\sub{Pl}=G\half=5.39\times 10^{-44}\sunit
\label{5}\ee
As we extrapolate back to this epoch, quantum gravity effects presumably
invalidate the concept of time, which conveniently removes the need to
discuss what is before the Big Bang!

\subsection{The matter content of the universe}

If gravity were negligible, $a(t)$ would increase linearly with time
and $\dot a$ would be constant. In fact, gravity slows down the
expansion making $\dot a$ a decreasing function of time.
To calculate the effect of gravity, consider a test particle
of unit mass, on the surface of a comoving sphere (one expanding with the
universe).
If $\rho$ is the mass density of the universe and $r$ is the
radius of the sphere, the mass inside it is $(4\pi/3)r^3\rho$
and the potential energy of the particle is
$-(4\pi/3) r^3 \rho G/r$. The kinetic energy of the particle is
$\dot r^2/2$ and therefore
\be
\dot r^2/2-(4\pi/3) r^3 \rho G/r=E
\label{6}\ee
where $E$ is the total energy of the particle.
Writing $r(t)=a(t)x$ and dividing through by
$a^2/2$ one arrives at the
{\it Friedmann equation}\footnote
{It is often stated that Newtonian gravity cannot be used to
discuss the expansion of the universe, and in particular that one cannot
justify the neglect of matter outside the sphere in the above
argument. This seems to me to be quite wrong. Drawing a much bigger
sphere around the one  that we considered, the matter within the bigger
sphere can certainly be neglected. There remains the matter outside the
sphere but it  accelerates the small sphere and the test particle equally
so it too can be ignored. This
is exactly the same argument that justifies the use
of Newtonian gravity for a galaxy or for the solar system, and
it seems to be just as good in the cosmological case.}
\be
H^2=\frac{8\pi G}{3}\rho -\frac{K}{a^2}
\label{7}\ee
where $K=-2E$. If $K\leq 0$ (energy $E\geq0$)
the expansion will continue indefinitely whereas if $K>0$
it will eventually give way to contraction leading to a
`Big Crunch'. The critical value separating these
possibilities is $K=0$, which corresponds to
Friedmann equation
\be
H^2=\frac{8\pi G}{3}\rho
\label{8}\ee
The corresponding mass density is called the
{\it critical density}, and its present value is
\bea
\rho_{c0}&=&1.88 \times 10^{-29} h^2 \mbox{\,g\,cm}\mthree\label{9}\\
&=&10.5 h^2 \GeV\munit\mthree\label{10}\\
&=&(3.0 \times 10^{-3}\eV)^4 h^2
\label{11}\eea
It is convenient to define the {\it density parameter} by
\be
\Omega=\rho/\rho_c
\label{12}\ee

Since mass is conserved, $\rho\propto a\mthree$.
Putting this into
the Friedman equation and remembering that $H=\dot a/a$, we have an
expression for $dt/da$ which can be integrated to give $t(a)$ and
hence $a(t)$. The result depends on
$K$, or equivalently on the present
density parameter $\Omega_0$. As I shall discuss in Section 6,
inflation strongly suggests the value $\Omega_0=1$,
and almost everyone working in the field believes that this is the true
value.\footnote
{In this context $\Omega$ includes all forms of energy density,
but in the matter dominated era other
forms of energy are supposed to be negligible
except conceivably for that due to a cosmological constant as discussed
later.}
In that case \eq{8} gives
$a=\propto t^{2/3}$.
Another case which is easy
to solve is $\Omega_0\ll 1$, which corresponds to negligible gravity
and $a\propto t$.

\subsubsection*{Baryonic matter}

What is the observed value of $\Omega_0$? Consider first the contribution
of ordinary matter {\it ie} nuclei and electrons. In the context of cosmology
this is usually called {\it baryonic matter} since the
baryons (nuclei) vastly outweigh the electrons. From
the nucleosynthesis calculation we know that the baryon contribution to
$\Omega_0$ is \cite{walker}
\be
\Omega_B = (.013\pm.002) h \mtwo <.09
\label{13}\ee
where I have used
$.4<h$ to obtain the upper limit. Thus if $\Omega_0=1$ there
exists {\it non-baryonic dark matter}, whose nature I discuss later.

The luminous matter in the universe, consisting of stars and
radiation-emitting gas,
accounts for only $\Omega_B\sim .01$. Unless $h$ is close to 1
(which we shall see in a moment is impossible with the favoured value
total value $\Omega_0=1$), it
follows from \eq{13} that there is a lot of {\it baryonic dark matter}.
If $\Omega_0=1$ it constitutes on average a few percent of the total amount
of dark matter. The fraction is presumably more or less the same
throughout inter-galactic space.
Within a given galaxy, one might expect the baryons to be concentrated
more in the central, luminous part than in the
dark halo. The reason is that
baryons
(ordinary matter) can emit radiation
whereas non-baryonic dark matter interacts too weakly to do so (or it
would not be dark). In consequence baryons can lose more energy,
allowing them to settle more deeply into the galaxy centre.
The baryons in a galaxy might be in the form of a non-emitting gas, or
they might be
failed stars with\footnote
{A collapsing gas cloud has to have mass $\gsim\msun$ to achieve a high
enough temperature for nuclear reactions to start, but on the other hand
very light gas clouds are thought not to collapse at all.}
 mass perhaps $\sim.001$ to $.1\msun$,
or dead stars (old white dwarfs, non-emitting
 neutron stars and the occasionally black hole arising from the collapse
of an exceptionally massive star) with mass $\sim\msun$.
These
objects are called MACHOS (massive compact halo objects),
because they will occur only in galaxy halos.
(As far as we know, bound objects form only within galaxies
so that the intergalactic baryons have to be in the form of
non-emitting gas.)

Detection of
MACHOS in our galaxy has recently been claimed through
their gravitational lensing of stars (microlensing),
a variation in brightness of the star occuring on a timescale of a few days
as a MACHO crosses the line of sight \cite{machos}.
If the observed MACHO density reflected the universal average
baryon dark matter density,
it would correspond to $\Omega_B\gsim .1$, but
since only the central part of the galaxy is probed this need not be the
case, and even a much higher MACHO mass density would be quite consistent
with the nucleosynthesis value $\Omega_B\simeq.05$.

\subsubsection*{Non-baryonic dark matter}

One can try to estimate the total amount of matter through its
gravitational effect. The gravitational field in a bound system
such as a galaxy or galaxy cluster can be deduced from the velocities of
its components, as evidenced by the Doppler effect (the `components'
can be gas molecules, stars or whole galaxies). One finds that
each galaxy is surrounded by a dark halo accounting
for most of its mass, the total galaxy contribution being
$\Omega_0\simeq .1$. Galaxy clusters contain in addition inter-galactic
gas, whose contribution to $\Omega_0$ is not yet known.

On larger scales, where the universe is almost homogeneous
and isotropic,
one can in a similar spirit observe the small departure from
uniform expansion. This defines a `peculiar velocity' field, which is
usually called the {\it bulk flow}. If one knew the bulk flow
$\bf v$
{\it and} the density perturbation $\delta\rho/\rho$,
one could deduce $\Omega_0$ through the relation (\cite{kotu},
cf. \eq{147})
\be
\frac{\del.{\bf v}}{3H_0}=-\frac13 (\Omega_0)^{.6}
\frac{\delta \rho}{\rho}
\label{14}\ee
To estimate these quantities one has to rely on observations of
the galaxies. Largely on the basis of numerical simulations of
gravitational collapse, it is generally assumed that their motion
accurately measures $\bf v$, but that their density contrast
is equal to a {\it bias factor} $b$ times the underlying density
contrast, with $b\sim1$ within a factor of 2 or so
(its value depends among other things
on whether one is looking at optical or infrared galaxies).
A recent study using this method \cite{dekel}
gives $(\Omega_0)^{.6}=1.28^{+.75}_{-.59} b_I$ where
$b_I$ is the bias factor for the infrared galaxies in the IRAS
catalogue. Including non-linear effects strengthens
this result slightly, so that for example if one assumes $b_I>.5$
one deduces $\Omega_0>.3$ at 95\% confidence level. A smaller bias
factor would certainly lead to problems with interpreting other data
so it seems fair to say that the bulk flow indicates
that $\Omega_0>.1$.
In view of the nucleosynthesis limit $\Omega_B<.1$ this means that
{\it non-baryonic dark matter seems to be needed}.

Further evidence about $\Omega_0$ comes from
what one might call the `standard model' of structure
formation. This model, which
has been the almost universal paradigm for
many decades and is by far the most thoroughly explored and successful
one,
supposes that the
structure originates as a  Gaussian adiabatic density perturbation, whose
spectrum at horizon entry is approximately scale independent.
It requires that $\Omega_0$ is at the upper end of the
range $0$ to $1$, with $\Omega_0=.1$ definitely excluded.

\subsubsection*{Summary}

{}From nucleosynthesis, baryonic matter contributes
$\Omega_0\simeq.01$ to $.09$.
The observed total $\Omega_0$, especially if one accepts the standard model
of structure formation, is higher, and the theoretically favoured
value is $\Omega_0=1$. Thus non-baryonic dark matter
is indicated. In addition, baryonic dark matter is indicated because
luminous matter accounts for only $\Omega_0\sim.01$.

\subsubsection*{Value of $\Omega$ at early times}

Even if $\Omega$ is not equal to 1 at the present epoch, it quickly
approaches 1 as we go back in time because the first term in \eq{7}
is proportional to $\rho\propto a\mthree$, and it therefore dominates
the second term which is proportional to $a\mtwo$. If $\Omega_0\ll 1$
we can easily estimate the epoch
before which $\Omega$ is practically equal to 1 as follows.
As long as $\Omega\ll 1$, gravity
is negligible and therefore $\dot a$ is constant, leading to $H\propto a$
and $\Omega\propto \rho/H^2\propto a\mone$.
If follows that $\Omega\simeq 1$ before the epoch
\be
a\simeq \Omega_0
\label{15}\ee
Since $\Omega_0\gsim .1$, we see that $\Omega$ is certainly close to 1
before $a\sim .1$.

\subsubsection*{The age of the universe}

One would like to define the
age $t$ of the universe at a given epoch as
the time since $a=0$ (in other words, to set $t=0$ at $a=0$).
As we noted earlier the Planck scale puts an absolute limit on the
earliest meaningful time, and more seriously
the matter dominated era that we are treating here goes back only
to $a\sim10\mfour$. So a practical definition of the age of the universe
at a given epoch is simply the time since $a$ was much less than the
current value.

The Friedmann equation allows one to calculate the present
age of the universe in terms of $\Omega_0$. For
$\Omega_0=1$ the behaviour $a\propto t^{2/3}$ gives at any epoch
\be
H=\frac2{3t}
\label{15a}
\ee
and therefore
\be
t_0=\frac23 H_0\mone=6.5\times 10^9 h\mone \yr
\label{16}\ee
The smallest conceivable value $\Omega_0\simeq.1$ gives
\be
t_0=.9 H_0\mone=8.8 \times 10^9 h\mone \yr
\label{17}\ee

An upper limit on the age of the universe is provided by
the age of the oldest stars (observed in globular clusters)
which is bigger than $1.0\times 10^{10}$\,years.
With the favoured value $\Omega_0=1$ this requires
$h<.65$, whereas with
$\Omega_0=.1$ the limit is $h<.88$.

\subsubsection*{The measured value of the Hubble parameter}

We noted earlier that using Hubble's law, different astronomers have
produced estimates of $h$ in the range $.4<h<1.0$. These Hubble law
estimates are generally agreed to be very difficult, whereas the
age limits that we have discussed are generally agreed to be sound.
If $\Omega_0=1$, and if there is no cosmological constant (see below),
one should therefore discard those Hubble law
estimates which conflict with the age bound. Since both
of these conditions are widely agreed to be likely, one concludes that
the current best guess for the true value of the Hubble parameter is
\be
.40 < h < .65
\label{18}\ee

\subsection{High redshift observations}

So far we have discussed only
the relatively nearby part of the universe, located at a distance small
compared with the Hubble distance. The redshift \eq{1} from galaxies in
this region can be written
\be
z=\diff a/a\ll1
\label{19}\ee
where $da$ is the change in the scale factor since the light was
emitted.
Thus, we see the nearby universe as it was
in the very recent past,
when the scale factor had almost its present value $a=1$.
Now we consider the distant universe, which is seen as it was in the
distant past.

In order to discuss the distant universe we need general relativity.
The central idea of GR, and practically the only one that I shall use in these
lectures, is
that all measurements are to be made locally. Thus, one populates the
universe with comoving observers. In the homogeneous, isotropic universe
comoving observers are in freefall (their worldlines are geodesics)
and as a result their clocks can be synchronized once and for all.
In other words, they all agree on the age of the universe $t$.
The hypersurfaces in spacetime with fixed $t$ are orthogonal to the
comoving worldlines, and will be called {\it comoving hypersurfaces}.

\subsubsection*{The redshift measures the scale factor}

To calculate the redshift from a distant object, emitting its light
when $a\ll 1$, we can add up the redshifts seen by a sequence of
comoving observers (observers moving with the expansion
of the universe), each separated by a distance $\ll H_0\mone$.
 In the small region around each one \eq{19} still applies except
that it is evaluated at the time $t$ when the radiation passes by.
Using $H=\dot a/a$ and integrating gives
\be
\frac{\lambda+\Delta \lambda}{\lambda}\equiv 1+z
=a\mone
\label{20}\ee
{\it The redshift measures directly the scale factor of the universe
at the time of emission of the radiation.}

\subsubsection*{The size of the observable universe}

We  would like to know the present distance of an object observed with
redshift $z$, defined as the sum of the present distances between
the comoving observers along its line of sight.
The distance $dr$ from one comoving observer to the next
is proportional to $a(t)$ and so can be written $dr=a dx$ where
$dx$ is independent of time. Thus each observer has a time independent
{\it comoving coordinate} $x$, related to his distance from us
by $r=ax$. Let us
work out the trajectory $r(t)=x(t) a (t)$ of a photon moving towards us.
During time $dt$ it passes between observers with coordinate distance
$x+dx$ and $x$, who are separated by distance $a \diff x=\diff t$ ({\it ie}
thes
 e
observers measure the photon speed to be 1). The comoving distance of the
source from us is therefore $x=\int^{t_0}_{t_e} \diff t/a$,
where
the subscript $0$ denotes the present and the subscript $e$ the time of
emission. Its
present distance is therefore
\be
r(t_0,t_e)=a_0\int^{t_0}_{t_e} \diff t/a
\label{20a}\ee
(It is convenient to
display the present value $a_0=1$
of the scale factor in this context.)
Given the time dependence of $a$,
 this relation gives the present distance of an
object observed with redshift $z$.

In the limit $z\gg1$ it gives the present distance that we can see in
principle, corresponding to the limit where the object emits its light
at $t=0$. This distance is called the {\it particle horizon} and is
given by
\be
r\sub{p.h.}(t_0)=
a_0\int^{t_0}_{0} \frac {\diff t}{a}
\label{21}\ee
Assuming that $\Omega_0=1$ one has
$a\propto t^{2/3}$ and therefore $r\sub{p.h.}(t_0)=2 H_0\mone$.

There is nothing special about the present, so we
can define the particle horizon $R\sub{p.h.}(t)$ at any epoch,
\be
r\sub{p.h.}(t)=
a(t)\int^{t}_0  \frac {\diff t}{a}
=a(t)\int^a_0 \frac1{aH}\frac{\diff a}{a}
\label{22}\ee
It gives the size of the biggest causally connected region, across which
a signal with the speed of light can travel since $t=0$.

As we noted earlier, the mathematical epoch $a=t=0$ has to be replaced
in practice by an epoch when $a$ and $t$ are much less than their current
 values. Thus, $r\sub{p.h.}(t)$ can really only be defined as $r(t,t_e)$
where $t_e$ is some early epoch.
As long as gravity is attractive,
$aH=\dot a$ increases as we go back in time, and from \eq{22}
$r(t,t_e)$ converges as $t_e\to 0$, and is typically
of order $H\mone(t)$ as we just found for our particular case ($\Omega_0=1$ and
matter dominating). But if the universe begins with an inflationary
era, where gravity is by definition repulsive,
$r(t,t_e)$ diverges. If we push $t_e$ back to the beginning of
inflation, it is much bigger than the Hubble distance
$H\mone(t)$, so a region much bigger than the Hubble distance is
causally connected.\footnote{At all epochs $t$ up to the present, on
the usual assumption that the observable
universe is inside the Hubble distance at the beginning of inflation.}
But to achieve this causal connection at a given epoch, one has to work with
signals emitted at a much earlier
epoch, and what usually matters physically is the size of a region that can be
causally connected by signals emitted in the last Hubble time or so,
namely the Hubble distance. In the context of the early universe,
the Hubble distance is usually referred to
as `the horizon', and has largely displaced the `particle horizon' as
the sort of horizon that cosmologists are interested in.

\subsubsection*{Non-Euclidean geometry?}

To go further in interpreting high redshift observations, one
needs to know the spatial geometry, {\it ie} the
geometry of the fixed $t$ hypersurfaces.
Einstein's field equation gives the spatial line element as
\be
\diff\ell^2=
\frac{\diff r^2}{1-Kr^2/a^2}+r^2(\diff\theta^2+\sin^2\theta\diff\phi^2)
\label{23}\ee
where $K$ is the quantity that appears in the Friedmann equation.
The distance element $dl$ is that measured by a local comoving observer.

A surface with fixed coordinate $r$ clearly has the geometry of a sphere
in Euclidean space with radius $r$, and in particular its area is
$4\pi r^2$. However the radial distance between adjacent spheres is
$(1-Kr^2/a^2)\mhalf\diff r$, which is equal to $dr$ only if $K=0$,
corresponding to $\Omega=1$. Thus {\it space is
non-Euclidean (`curved') if $\Omega\neq1$.}
If $\Omega<1$ space is infinite just as in the Euclidean case, but if
$\Omega>1$ it has the remarkable property of being finite but unbounded.
In particular, a surface made out of geodesics, such as the
`equatorial plane' $\phi=\pi/2$, has the geometry of a sphere
whose curvature is $R^2=K/a^2$. The quantity $R$ is the
radius that the sphere would have if it lived in Euclidean space,
or equivalently it is defined by saying that the area of the sphere is
$4\pi R^2$.
The total volume
of space, obtained by integrating the volume $4\pi r^2 (1-Kr^2/a^2)
\mhalf dr$ between adjacent spheres, is $16\pi^2 R^3$.

One calls $K/a^2$ the `curvature of space' whether it is positive
($\Omega>1$) or negative ($\Omega<1$).
The departure from Euclidean
geometry becomes significant when
$r^2\gsim |a^2/K|$, and one refers to
the distance $|a^2/K|\half$ as the
{\it curvature scale}.
 If  $\Omega\simeq 1$ the curvature scale is
much bigger than the Hubble distance $H\mone$,
if $|1-\Omega|\ll 1$ it is of order $H\mone$, and if
$\Omega\gg1$ it is much less than $H\mone$.

\subsubsection*{High redshift observations}

Having displayed the basic tools I will briefly say what we learn from
observation of the distant universe.
The most important finding is that the universe is definitely
evolving. The most dramatic case is that of
quasars (active galactic nuclei), whose abundance per comoving volume
peaks at $z\sim 3$ or so. Neither quasars nor
any other objects are observed
at $z\gsim 5$. Ordinary galaxies as well as clusters
are observed out to a redshift of order $1$ to $2$, and they too show
signs of evolution. (The mere fact that galaxies have size of order
one tenth of their spacing and do not expand with the universe means, of
course, that they cannot have existed before $z\sim10$.)

Ideally, high redshift observations plus an understanding of galactic
evolution would give us information the value of $\Omega_0$.
In practice the information obtained in this way does not add anything
to our knowledge as yet,
except for the issue of a cosmological constant
that we look at next.

\subsubsection*{The Robertson-Walker metric}

Before leaving the description of the large scale universe I should
write down the famous {\it Robertson-Walker spacetime metric},
\be
\diff s^2=\diff t^2-
a^2(t)\left[
\frac{\diff x^2}{1-Kx^2}+x^2(\diff\theta^2+\sin^2\theta\diff\phi^2)
\right]
\label{24}\ee
It expresses the fact that the universe is homogeneous and isotropic,
with spatial geometry given by \eq{23}.
The spacetime metric is
the traditional starting point for cosmology but in these lectures
I will consistently avoid it, in favour of the
physical notion of comoving worldlines.

\subsection{A cosmological constant?}

The discussion so far assumes that a) gravity is described by Einstein's
field equation, and b) the only significant source of gravity is
the matter. People have considered the possibility of
alter assumption a) by adding a {\it cosmological constant} $\Lambda$ to the
lagrangian of Einstein gravity. This is equivalent to
to keeping Einstein gravity, but modifying assumption b) by invoking an
additional source of gravity in the form of a fluid with pressure $p_\Lambda$
and energy density $\rho_\Lambda$ related by $p=-\rho$.
This second viewpoint is the most useful and I adopt it here.
{}From
\eq{50} below it then follows that $\rho_\Lambda$ is time independent,
so we are considering the possibility that
\be
\rho(t)=\rho_m(t)+\rho_\Lambda
\label{25}\ee
The corresponding present density parameter is
\be
\Omega_0=\Omega_m+\Omega_\Lambda
\label{26}\ee

If we are ever faced with a measured value $\Omega_m<1$,
we will probably want to invoke a cosmological constant
 to
keep the total value $\Omega_0=1$ which inflation suggests.
A cosmological constant is, however  unattractive from
a particle physics viewpoint because the
corresponding tiny vacuum energy density
$(.001eV)^4$ (\eq{11}) is difficult to understand.
What about observation? Assuming that the total
$\Omega_0=1$, the present the situation is as follows
\cite{carroll,maoz}.

On scales much less than the Hubble distance the effect of a
cosmological constant is to add a repulsive term to Newtonian
gravity, but the extra term is insignificant.

Going on to observations of the distant universe,
gravitational lensing statistics give a remarkably good limit
$\Omega_\Lambda\lsim.6$ \cite{maoz},
and structure formation gives a similar but
less reliable limit. Thus $\Omega_m>.4$, which means that one cannot use
a cosmological constant to avoid the need for non-baryonic dark matter.

Coming to the age of the universe, \eq{52} shows that $\Omega_\Lambda$
corresponds to repulsive gravity, $\ddot a>0$, so the age of the
universe will be an increasing function of $\Omega_\Lambda$.
In fact, the upper limit on $h$
increased from $.65$ to $.88$ as $\Omega_\Lambda$ increases from
$0$ to $.6$. Thus a high value of $h$ measured by, say, Hubble's law
would be evidence in favour of
a cosmological constant, and conversely a confirmation of the expected
low value would be evidence
against it.

In summary, a cosmological constant is theoretically objectionable,
and is not so far required by observation.

\section{The Radiation Dominated Era}

\subsection{Overview}

An epoch in the early universe is conveniently specified by giving
the temperature $T$ in eV, MeV or GeV (as usual in cosmology I set $k_B=1$).
The epoch from $1\MeV$ to $1\eV$, which can be taken to mark the end of
the early universe, is well understood,
and is highly
constrained because it includes the epoch of nucleosynthesis.
At earlier epochs one encounters a lack of knowledge about
the nature of the quark-hadron and electroweak phase transitions,
and eventually a basic ignorance of the
fundamental interactions and the initial conditions.
Later one enters the matter dominated era, when structure formation
occurs and again our knowledge becomes less secure. The various epochs
are summarised in Table 1.

\subsection{The cosmic microwave background (cmb)}

The most direct evidence for a hot early universe comes from the
cosmic microwave background (cmb). It is extremely isotropic, with a
very accurate blackbody distribution corresponding to temperature
$T=(2.736\pm.017)^0$\,K.

To understand what the cmb implies, recall that the blackbody distribution
specifies the photon occupation number of each momentum state as
\be
f(p) = 2[e^{(p/T)}-1]\mone
\label{27}\ee
The factor 2 counts the number of
spin states, and $p$ is the momentum. Since there are
$(2\pi)\mthree \diff^3 p\diff^3 x$ states  in a
given volume of phase space, the number density $n$ and the energy
density $\rho$ of the photons are given by
\bea
n&=&\frac{2}{(2\pi)^3} \int^\infty_0 f(p) 4\pi p^2 \diff p
\label{28}\\
\rho&=&\frac{2}{(2\pi)^3} \int^\infty_0 E f(p)  4\pi p^2 \diff p
\label{29}\eea
with $E=p$. Evaluating the integrals gives
\bea
\rho_\gamma&=&2(\pi^2/30)  T^4
\label{30}\\
n_\gamma&=& 2(\zeta (3)/\pi^2)  T^3
\label{31}\eea
where $\zeta(3)=1.2021$.
The photon energy density is negligible compared with the mass density
of the matter, corresponding to present density parameter
\be
\Omega_\gamma\equiv\rho_\gamma /\rho_{c0}=2.51\times 10^{-5} h\mtwo
\label{32}\ee
Dividing $n$ by the baryon number density
$n_B=11.2 h^2 \munit\mthree \Omega_B$,
\be
\eta\equiv\frac{n_B}{n_\gamma}=2.68\times 10^{-8} h^2\Omega_B
=(3.4\pm.6)\times 10^{-10}
\label{33}\ee
(The last equality is the nucleosynthesis result \eq{13}.)
Thus there are many photons per baryon.

Now let us evolve the blackbody distribution back in time,
assuming that the
number of photons in a comoving volume is conserved
(as we shall see this is correct back to the epoch of electron-positron
annihilation). Because of the redshift, the momentum
of each photon goes like $a\mone$, which is equivalent
to saying that the blackbody distribution is preserved, with $T\propto
a\mone$.
Since $T\propto a\mone$,
$\rho_\gamma\propto a\mfour$, so $\Omega_\gamma/\Omega_m
\propto a\mone$. We conclude that
{\it before the epoch $a\mone\sim 10^4$ the energy density of the
cmb photons dominates that of the matter.}

\begin{table}
\centering
\begin{tabular}{|l|l|l|}
\hline \hline
TEMPERATURE  & EVENT & CONSEQUENCES \\
\hline
$\mpl=10^{19}\GeV$ & Inflation begins? & Immediate collapse avoided\\
\hline
$10^{16}$--$10^{11}\GeV$? & Cosmological scales &  Field fluctuations\\
& leave the horizon & become classical\\
\hline
$10^{16}$--$10^{11}\GeV$? & Inflation ends & Particles created \\
\hline
$10^{16}\GeV$--$1\TeV$?? & Particles thermalize & Radiation domination
begins \\
\hline
$1\TeV$-$100\GeV$ & Electroweak phase & Particles acquire mass \\
        & transition & Baryogenesis occurs? \\
\hline
$100\MeV$ & Chiral and quark/hadron & Pions acquire mass and\\
 & phase transitions & quarks bind into hadrons\\
\hline
$1\MeV$ & Neutrinos decouple & Neutrons start to decay \\
\hline
$1\MeV$ & $\e\ebar$ annihilation & Afterwards $(T_\gamma/T_\nu)=
(11/4)^{1/3}$ \\
\hline
$.1\MeV$ & Nucleosynthesis & Most neutrons bind into $\Hefour$ \\
\hline
$1\eV$ & Matter domination & The cold dark matter density \\
 & begins & contrast starts to grow\\
\hline
$.3\eV$ & Atoms form & Photons decouple and the baryonic\\
 & & matter density contrast grows\\
\hline
$\sim 10\mthree\eV$? & First galaxies form & \\
\hline
$2.4\times 10\mfour\eV$ & The present epoch & \\
\hline \hline
\end{tabular}
\caption[The early universe]{The early universe.
The left hand column defines the epoch. It gives the temperature,
which is roughly equal to $\rho\quarter$ where $\rho$ is the energy
density (in the first three rows the universe is not actually thermalised
so the number in the left hand column is {\it defined} as
$\rho\quarter$). During radiation domination the
time since the big bang is roughly $1\MeV/T$ seconds, and
during both epochs the scale factor is roughly $a\simeq T_0/T$ where
$T_0\simeq10\mfour\eV$ is the present temperature of the cosmic
microwave background.}
\end{table}

\subsection{Physics in the radiation dominated era}

Motivated by the observed cmb, one makes the hypothesis that the
early universe is a hot ideal gas. The hypothesis is found to be
consistent in the context of gauge theories like the Standard Model.
It leads to the prediction that the {\it nucleosynthesis}
of $\Hefour$ and a few other light nuclei
occurs in the early universe, and the calculated primordial abundances
are in striking agreement with observation.

\subsubsection*{Adiabatic expansion}

The pressure of a gas is given by $p=v^2 \rho/3$, where $v^2$ is the
mean square particle velocity and $\rho$ is the energy density.
 It follows that
\bea
p&=& \rho/3\mbox{\hspace{5em}relativistic}
\label{41}\\
p&\simeq& 0\mbox{\hspace{5em}nonrelativistic}
\label{42}\eea
As the universe expands, the change in the
energy $E=a^3\rho$ in a comoving volume
$a^3$ is given by $dE=-p d(a^3)$ so that
\be
a\frac{\diff \rho}{\diff a}=-3(\rho+p)
\label{43}\ee
For a nonrelativistic gas this gives $\rho\propto a\mthree$,
but for a relativistic gas it gives
\be
\rho\propto a\mfour
\label{44}\ee
The extra factor $a\mone$ arises from the redshift of the
particle energies between collisions, each collision conserving
energy.

The homogeneity and isotropy of the early universe implies that its
expansion is adiabatic (no heat flow), so that entropy is conserved.
Keeping only relativistic particles in the gbb, the entropy
density is
\be
s=\frac{\rho+p}{T}=\frac43\frac{\rho}{T}=\frac43\frac{\pi^2}{30}g_* T^3
\label{45}\ee
Conservation of the
entropy $a^3 s$ in a comoving volume therefore gives the
{\it relation between the temperature $T$ and the scale factor
$a$}
\be
aT\propto g_*\mthird
\label{46}\ee

For rough estimates one can ignore the time dependence of $g_*$
to deduce that $T\propto a\mone$, and can also set $s\sim n$ to
deduce that {\it particle number is roughly conserved}.
In particular,
\eq{33} holds at least roughly in the early universe,
\be
n_B/n_\gamma\sim 10^{-10}
\label{47}\ee
For a detailed calculation one has to take into account that $aT$
increases abruptly whenever the temperature falls below the mass
of a particle species. Physically, the relative increase in
temperature comes from radiation emitted when the
species annihilates and/or decays.

\subsubsection*{The Friedmann equation}

According to general relativity, the Friedmann equation \eq{7}
that we derived from Newtonian gravity remains valid, if
$\rho$ denotes the energy density instead of just the mass density.
One can differentiate it using \eq{43}, which
is equivalent to
\be
\dot\rho=-3H(\rho+p)
\label{50}\ee
The result is
\be
\dot H=-H^2-\frac{4\pi G}{3} (\rho+3p)
\label{51}\ee
or
\be
\frac{\ddot a}{a}= -\frac{4\pi G}{3} (\rho+3p)
\label{52}\ee
This last expression reduces to the Newtonian result in the
limit $p\ll\rho$, if $\rho$ is identified with the mass density.
But in general both energy {\it and pressure} are sources of gravity.

Returning to the Friedmann equation, the density parameter
$\Omega$ is still defined by \eq{12}. Setting it equal to 1,
corresponding to $K=0$, and remembering that for radiation
$\rho\propto a\mfour$, one finds
\be
a\propto t\half
\label{53}\ee
corresponding to
\be
H=\frac1{2t}
\label{54}\ee

\subsubsection*{Thermal equilibrium}

In the early universe collision and decay processes are continually
creating and destroying particles. Let us suppose that it is in
thermal equilibrium, which means that each process is taking place
at the same rate as its inverse. Then the
the number of particles of a given species, per momentum state, is given
by
\be
f(p)=g[e^{(E-\mu)/T}\pm 1]\mone
\label{34}\ee
In this expression, $g$ is the number of spin states, $p$ is the
momentum, and $E$ is the energy, given in terms of the mass $m$ by
$E=\sqrt{p^2+m^2}$. The sign is $+$ for fermions
(ensuring that there is at most one particle per quantum state),
and $-$ for bosons.
The quantity $\mu$ (which depends on temperature) is called
the chemical potential of the species. Given the chemical potential,
this expression
determines the number density $n$ and energy density $\rho$ of the
species through \eqs{28}{29}.

The chemical potential
is conserved in every collision, so that if a process $\A\B\to\C\D$
occurs then $\mu_A+\mu_B=\mu_C+\mu_D$.
In the early universe all known particle species
are freely created and destroyed (provided that their mass does not
exceed the typical collision energy, which we shall see in a moment is
of order $T$). The only significant restriction is
that each collision must respect the conservation of any
`charges' that the particles carry. In the context of the Standard Model
there are five such charges, namely electric charge, baryon number and the
three lepton numbers.\footnote
{The non-perturbative violation of the last four is not relevant in this
context.}
Since the photon carries none of these `charges',
a single photon can be created through processes of the
form $\A\to\A\gamma$, which implies that $\mu=0$ for the photon,
leading to the blackbody
distribution. The same goes for any particle which is its own
antiparticle, so it too has $\mu=0$. If the antiparticle is distinct
one can create a particle-antiparticle pair
through some process of the form $\A\to\A+$pair,
which implies that the particle and its antiparticle have opposite
$\mu$. As a result, $\mu$ vanishes
if the number densities $n$ and $\bar n$ of
particle and antiparticle are equal, and otherwise
is determined by the imbalance  $n-\bar n$.

More generally, one can show that until weak interaction equilibrium fails
at $T\sim 1\MeV$, enough reactions occur to determine all of
the chemical potentials
in terms of the densities of the five conserved
`charges'. Furthermore, all of the chemical potentials are zero
if the `charges' are all zero.
In that case one
has what I shall call the {\it
generalized blackbody distribution} or gbb
\be
f(p)=g[e^{E/T}\pm 1]\mone
\label{35}\ee
Putting this into \eqs{28}{29} gives $n$ and $\rho$.
If the temperature is well above the mass of the species in question,
$m\ll T$, then one is in {\it relativistic regime}
and it is a good
approximation to set $E=p$. Using \eqs{28}{29} gives
\bea
\rho&=&\left\{ \begin{array}{rl}
     (\pi^2/30) g T^4 & \hspace*{5mm}\mbox{ bosons} \\
(7/8)(\pi^2/30) g T^4 & \hspace*{5mm}\mbox{fermions}
\end{array} \right.
\label{36}\\
n&=& \left\{ \begin{array}{rl}
(\zeta (3)/\pi^2) g T^3 & \hspace*{5mm}\mbox{ bosons} \\
(3/4)(\zeta (3)/\pi^2) g T^3 & \hspace*{5mm}\mbox{fermions}
\end{array} \right.
\label{37}\eea
Thus {\it according to the gbb, each relativistic species
contributes $\sim T^4$ to $\rho$ and $\sim T^3$ to $n$}.
(Note that the typical energy $E\sim\rho/n$ is of order $T$, so this is indeed
the relativistic regime.)
As $T$ falls below $m$ one moves
into the nonrelativistic regime, and according to the gbb
$\rho$ and $n$ fall rapidly (like $e^{-m/T}$).
The physical reason is that the typical energy $E$ available in a
collision is now insufficient to create the species.

Is the gbb valid in the early universe?
The charge density of the universe is certainly zero to very high
accuracy, or the expansion of the universe would be governed by
electrical repulsion instead of gravity.
The baryon number is not zero, but it
is small in the sense that
\be
\eta\equiv n_B/n_\gamma\ll 1
\label{38}\ee
The three lepton number densities cannot be measured directly,
 but let us suppose that
they too are small in the same sense.
It can be shown that if this is so,
the gbb is valid to high accuracy
for all relativistic species, the reason being that
for each species one then has
$|n-\bar n|\ll n_\gamma\sim T^3$. From now on I assume that the gbb is
valid for all
relativistic species in equilibrium, and the
success of the nucleosynthesis
calculation will  support that assumption.

As the temperature falls below the mass $m$ of a given species,
particle-antiparticle pairs at first rapidly annihilate
as dictated by the gbb.
Then the small particle-antiparticle imbalance becomes significant, and
annihilation soon stops because only particles
(or only antiparticles) survive.
Even if they do not decay, the
surviving particles give a negligible
contribution to
$\rho$ during radiation domination, and an even more negligible
contribution to $n$.

As long as a non-relativistic particle species is in equilibrium,
it follows from \eqs{34}{28} that
its number density is
\be
n=g\rfrac{mT}{2\pi}^{3/2} e^{-(m-\mu)/T}
\label{40a}\ee
We shall use this equation to determine the relative numbers of
different species of nuclei in thermal equilibrium, and the degree of
ionization of a gas in thermal equilibrium.

The upshot of this discussion is
that unless
one happens to be
at an epoch when $T$ is close to one of the particle masses,
$\rho$ and $n$ come just from the
relativistic species. In particular,
\be
\rho=(\pi^2/30) g_*(T) T^4
\label{39}\ee
where the {\it effective number of degrees of freedom}
$g_*$ is given by
\be
g_*(T)=\sum\sub{bosons} g+\frac78 \sum\sub{fermions} g
\label{40}\ee
where the sums go over particle species with $m<T$.
According to the Standard Model it
is of order 10 for $1\MeV\lsim T\lsim 10\MeV$ and of order
$100$ for $T\gsim 1\GeV$.
Extensions of
the Standard Model, notably supersymmetry, introduce new particles
which can increase $g_*$ by a factor of a few at $T\gsim100\GeV$,
but usually leave it unchanged at lower energies.

\eq{4} applies only if all relativistic species are in thermal equilibrium.
This is true in
the Standard Model (for $T\gsim 1\MeV$), but not necessarily in extensions of
it.

\subsubsection*{The timescale}

Using \eqss{39}{8}{54}, the
timescale during radiation domination is
\be
\frac t{1\sunit}=2.42 g_*\mhalf \rfrac{1\MeV}{T}^2
\label{55}\ee

\subsection{The Standard Model of the early universe}

\subsubsection*{Interaction rates and the ideal gas condition}

Our basic assumptions are that the early universe is an ideal gas,
and that it is in thermal equilibrium. The first requires that
the interactions are not too strong, but the second requires that
they are strong enough to make the relevant reaction rates
(per particle) be bigger than the rate of expansion $H$ of the universe.
Which reactions are `relevant' depends on what one supposes about the
chemical potentials; each reaction that is in equilibrium
implies a relation between the
chemical potentials (if it creates or destroys particle species) and
there have to be enough relations to ensure the postulated conditions on
the chemical potential. In particular, if we want the gbb to hold for
relativistic particle species, enough interactions have to be in
thermal equilibrium to ensure that all of the chemical potentials vanish
(on the assumption that the `charge' densities all vanish). In the
context of the standard model, one can show that this is indeed ensured
if all processes of the form $\A\B\to \C\D$
going by the exchange of a single gauge boson are in equilibrium.
Let us check that such reactions are indeed in thermal equilibrium, and
that the interactions are yet weak
enough to ensure the ideal gas condition.

Suppose first that the gauge boson is massless; this covers the
case of the electromagnetic interaction before photon atoms form
(photon decoupling) the strong interaction before
hadrons form (the QCD phase transition) and the weak interaction
before the W and Z acquire mass (the electroweak phase transition).
For an isolated collision the cross section $\sigma$
is infinite because the interaction has
infinite range, but in a gas there is a cutoff at
the particle spacing $n\mthird$ (Debye shielding).
Using $n\sim T^3$ one finds that $\sigma\sim \alpha^2/T^2$,
where $\alpha$ is the electromagnetic, weak or strong
coupling strength as appropriate.
If at least one of the initial particles is relativistic,
the rate per particle is
$\Gamma\sim n\sigma$ (the formula is exact if one
particle is at rest, and a factor 2 too small for a head-on relativistic
collision.)
Using \eqs{54}{55}
 this gives $\Gamma/H \sim .1 \alpha^2 \mpl/T$. Taking the high
energy estimate $\alpha\sim .1$, this is indeed
$\gg 1$ unless $T\gsim 10^{16}\GeV$. Thus the interaction is strong
enough to maintain thermal equilibrium, except at these very high
temperatures. What about the ideal gas condition? We can take it to be
the requirement that the mean free path $r$, determined for relativistic
particles by
$r\sigma n\sim 1$, is big compared with the
particle spacing $n\mthird$. Thus we require $n\twothird\sigma
\ll 1$ which is equivalent to $\alpha\ll 1$.
This is satisfied, except for the strong interaction
just before the QCD transition.

Now consider the weak interaction after the
electroweak phase transition. The
cross section is
$\sigma\sim \alpha^2 T^2/m_W^4$, which gives $\Gamma/H
\sim \mpl \alpha^2 T^3/m_W^4$. Taking the weak interaction
coupling as $\alpha\sim.01$ this is equal to 1 when
\be
T\sim \alpha^{-2/3}(m_W^4/\mpl)\third\sim 1\mbox{\ to }10\MeV
\label{56}\ee
Weak interactions are in equilibrium only above this temperature.
Since the cross section is
a factor $(T/m_W)^4$ smaller than in the previous case, the ideal gas
condition is amply satisfied for the weak interaction.

\subsection{The history of the early universe}

\subsubsection*{The electroweak phase transition}

The electroweak symmetry of the standard model is restored
above some critical temperature $T_c \sim 100\GeV$. The
Higgs field is then zero, corresponding to a false vacuum
with energy density  $\sim (100\GeV)^4$, and all particle masses vanish.
As the temperature falls through $T_c$ the Higgs field rapidly settles
to its true vacuum value and particles acquire mass.

The consequences of a phase transition can be very different, depending
on whether it is of first order or second order. In the first case
the transition is relatively violent because
bubbles of the new phase are produced. In the second case it is more
gently, the new phase being produced smoothly everywhere
(except near topological defects, which are not produced at the
electroweak phase transition).

It is not clear if the electroweak transition is of first or second
order, and the answer in any case should probably take into account
an extension of the Standard Model. If it is of first order
it will violate
baryon- and lepton number conservation because of
non-perturbative effects occurring in bubble walls.
As Shaposhnikov explains in his lectures, such a mechanism
might be responsible for generating the observed ratio
$n_b/n_\gamma\sim 10^{-10}$, either within the standard model or within
some extension of it.

\subsubsection*{The quark-hadron transition}

As the temperature falls through
$\Lambda\sub{QCD}\sim100\MeV$, chiral symmetry is spontaneously broken which,
among other things, gives mass to the pion. More importantly,
the {\it quark-hadron} phase transition occurs, which binds
quarks and gluons into hadrons as their typical spacing
$\sim T\mone$ becomes bigger than
$\Lambda\sub{QCD}\mone\sim 1$\,fm.

The quark-hadron phase transition \cite{silvio} is thought to be a second order
one. If it turned out to be first order though, the bubbles could lead
to dramatic effects, perhaps interfering with nucleosynthesis or even
creating black holes accounting for the dark matter. The mass of
such black holes might be of order $1\msun$ corresponding to
\eq{68} below, or much smaller.

\subsubsection*{Neutrino decoupling and $\e\ebar$ annihilation}

At $1\MeV\lsim T \lsim 100\MeV$, the only particles present
(according to the Standard Model) are photons, free protons and neutrons,
electrons, positrons, and the three neutrino species plus
their antiparticles. Other elementary particles as well as hadrons
are absent because they are unstable, and are too heavy to be created in a
typical collision ($m\gg T$). Nuclei, not to mention atoms and molecules,
are absent because their binding energy is much less than $T$, so
that collisions would destroy them.

The protons and neutrons
turn into each other through reactions like
$\e\p \leftrightarrow\nu_e \n$, $\ebar\n \leftrightarrow\bar\nu_e \p
$
and $\n \leftrightarrow\p\e\bar\nu_e$. As a result the ratio
$n/p$ of their number densities is determined by thermal equilibrium.
The chemical potentials satisfy
$\mu_p-\mu_n=\mu_{\nu_{\mbox{\tiny e}}}- \mu_e$,
with $\mu_{\nu_{\mbox{\tiny e}}}$ and $\mu_e$ both $\ll T$ (on the
assumption that the lepton number densities are sufficiently small
that the $\e$ and $\nu_e$ satisfy the gbb to good accuracy).
{}From \eq{40a} with $|\mu_p-\mu_n|\ll T$,
it follows that the neutron to proton ratio is
\be
n/p=e^{-\Delta m/T}
\label{57}\ee
where $\Delta m\equiv m_n-m_p=1.3 \MeV$.

When the temperature falls through $1\MeV$, three essentially unrelated
things happen to occur at about the same time.
\begin{itemize}
\item Neutrinos decouple because of \eq{56}
\item Electron-positron annihilation occurs because $m_e\sim 1\MeV$
\item Neutron creation stops because of \eq{56}.\footnote
{It would stop anyway, because to create a neutron
requires a collision energy bigger than
$m_n-m_p=1.3\MeV$.}
\end{itemize}
A detailed calculation shows that the events actually occur in the
indicated order; let us discuss them in turn.

After decoupling the neutrinos travel freely, maintaining their
gbb with temperature $T_\nu\propto a\mone$.
The photon temperature, though, is relatively raised by $\e\ebar$
annihilation. Before annihilation there are electrons, positrons and
photons in equilibrium giving
$g_*=11/2$, whereas after there are only photons giving
$g_*=2$. From \eq{46} it follows that after annihilation,
$T_\gamma=(11/4)^{1/3}T_\nu$.
Remembering that there are two spin states for the
photon and one for each massless neutrino or antineutrino species,
\eq{37} gives
\be
n_\nu=(3/11)n_\gamma
\label{48a}\ee
for each neutrino
species. If
all three neutrino species are massless, \eqs{31}{37} give
the neutrino contribution to $\Omega_0$,
\be
\Omega_\nu=(21/8)(4/11)^{4/3}\Omega_\gamma
=.681\Omega_\gamma
\ee
Using \eq{32} the epoch of matter-radiation equality is
therefore
\be
z\sub{eq}=2.37\times 10^4 h^2
\label{49}\ee

After neutrons cease to be created,
the np ratio is frozen in except for the slow
decrease caused by neutron decay. Its initial value
is $n/p\simeq1/6$, and by the time
that nucleosynthesis occurs at $T\simeq.1\MeV$ it has fallen to
\be
n/p\simeq 1/7
\label{58}\ee

\subsubsection*{Nucleosynthesis}

A nucleus is defined by the number $Z$ of protons, and the number
$A-Z$ of neutrons. If strong nuclear reactions occur at a sufficiently
high rate, while the weak interaction changing
protons to neutrons is negligible, the
chemical
potential of a given nucleus satisfies $\mu=Z\mu_p+(A-Z)\mu_n$.
By virtue of this relation, the $\mu$'s cancel in the ratio
$n_p^Z n_n^{A-Z}/n$ given by \eq{40a},
leading to the following number density for
a given nucleus
\be
n=g A\threehalf 2^{-A} \rfrac{2\pi}{m_N T}^{3 (A-1)/2}
n_p^Z n_n^{A-Z} e^{B/T}
\label{59}\ee
where $B$ is its binding energy. Together with the np ratio
and the total baryon density $n_B$ this determines all of the
nuclear densities. Using $n_B=\eta\n_\gamma$ together with
\eq{31} for the photon number density $n_\gamma$, one can
calculate the mass fraction in a given nucleus,
\bea
X&=&g[\zeta(3)^{A-1}\pi^{(1-A)/2}2^{(3A-5)/2}]A^{5/2}\eta^{A-1}
\nonumber\\
&\times& X_p^Z X_n^{A-Z} \rfrac{T}{m_N}^{3(A-1)/2} e^{B/T}
\label{61}\eea
If the baryon to photon ratio $\eta$ were of order 1, this expression
would say that the nuclei form  ($X$ becomes of order 1)
at roughly $T\sim B$, which is the
epoch after which a {\it typical} photon has too little energy to
destroy the nucleus. But because there are $\eta\mone\sim 10^{10}$
photons per baryon, the nuclei do not form until $T$ is well
below $B$.  The reason is that until that happens
there are still a lot of exceptionally energetic
photons around with energy of order $B$.

To determine whether a given set of nuclei will actually be in thermal
equilibrium at a given epoch in the early universe, one has to
take the rates of the relevant nuclear reactions from accelerator
experiments. (They cannot all be calculated accurately
from first principles, and are certainly not amenable to the order of
magnitude arguments that work for elementary particles.)
One finds that at
$T\gsim .3\MeV$ the lightest few nuclei are indeed in thermal
equilibrium, but that their number densities are negligible.
In the range $.1\MeV\lsim T\lsim .3\MeV$, thermal equilibrium
predicts that practically all of the neutrons are bound into
$\Hefour$, but when $T$ first enters this range thermal equilibrium
actually
fails badly because the reactions forming $\Hefour$ do not occur fast
enough to keep up with the `demand' for $\Hefour$
(the deuterium bottleneck). As a result most of the
neutrons bind into $\Hefour$ at $T\sim .1\MeV$, rather than at
$T\simeq .3\MeV$ as nuclear equilibrium would predict.
Assuming that all of the neutrons bind into $\Hefour$, its abundance by
weight is
\be
X=\frac{2n}{n+p}
\label{62}\ee
Taking $n/p\simeq1/7$ gives $X\simeq 22\%$.

The assumption of thermal equilibrium gives a reasonable
estimate of the $\Hefour$ abundance, but an accurate calculation of it
and even an estimate of the other abundances requires an out of
equilibrium calculation. This has to be done on a computer, taking the
rates for the various nuclear reactions from laboratory experiments.
The latest results from such calculations \cite{walker}
are shown in Figure 1 as a function
of $\eta$.

\begin{figure}
\vspace*{10cm}
\caption[Nucleosynthesis predictions]
{Nucleosynthesis predictions and observations, taken from
\cite{walker}. The upper section gives the $\Hefour$
abundance by weight, and the lower section gives the other abundances by
number. The horizontal lines indicate the observed values;
from top down the upper and lower limits for $\Hefour$, the upper limit
for $\D+\Hethree$, the lower limit for $\Hethree$ and the upper and lower
limits for $\Liseven$. The horizontal axis is $10^{10}\eta$,
and consistency with all observations requires that it be in the
shaded region.}
\end{figure}

The primordial abundances of $\Hefour$ and other light nuclei can be
measured by a variety of techniques, and the predicted
abundances agree provided that $\eta$ is in the interval indicated,
which corresponds to the result already quoted in \eq{33}.
The agreement is extremely impressive, and constitutes one of the
strongest reasons for believing the hypothesis that went into the
calculation. It also limits greatly ones freedom to alter them;
in particular it severely constrains possible
extensions of the Standard Model.

\subsubsection*{Photon decoupling}

After nucleosynthesis there are photons, protons, $\Hefour$ nuclei and
electrons in
thermal equilibrium. When the temperature falls sufficiently, most of
the electrons bind into hydrogen atoms, and one can use the thermal
equilibrium fraction of hydrogen to determine when this happens.
The atoms form through the process $\p+\e\to \H+\gamma$,
so that $\mu_p+\mu_e=\mu_H$.
The chemical potentials cancel in the ratio
$n_p n_e/n_H$ given by \eq{40a}, leading to
\be
n_H=n_p n_e \rfrac{m_e T}{2\pi}\mthreehalf e^{B/T}
\label{63}\ee
Here $B=m_p+m_e-m_H=13.6\eV$ is the binding energy of
hydrogen, and I have used
$g_p=g_e=2$ and $g_H=4$.
Ignoring the $\Hefour$ for simplicity,
$n_e=n_p$ and $n_B=n_p+n_H$,
 which with \eq{31} determines the
fractional ionization $X\equiv n_p/n_B$,
\be
\frac{1-X}{X^2}=\frac{4\sqrt2\zeta(3)}{\sqrt\pi}\eta
\rfrac{T}{m_e}\threehalf e^{B/T}
\label{64}\ee
This is called the {\it Saha equation}.
As with nuclei, very few atoms form until $T$ is well below
$B$, because there are so many photons per baryon.
In fact, one finds that $T\simeq.3\eV$ when $X$ has fallen to
$10\%$. The corresponding redshift is $z\simeq 1300$.
Soon after this, at $z\simeq 1100$ to $1200$, the photons decouple.

\subsection{Beyond the Standard Model}

There are several theoretical
reasons for wanting to extend the Standard Model.
\begin{itemize}
\item
One-loop contributions to the Higgs mass have to be cut off
at the scale $100\GeV$ to $1\TeV$, or they will be too big.
To impose the cutoff one can either make the Higgs composite,
as in technicolour type models, or invoke supersymmetry.
Composite models are becoming increasingly unviable in the light of the
ever more impressive agreement between the Standard Model and collider
experiments, notably at CERN, and I will not consider them here.
Supersymmetry requires that each known particle species has an as yet
undiscovered partner. The lightest supersymmetric partner is typically
weakly interacting and stable, and
is therefore a dark matter candidate.
\item
Unless one of the quark masses vanishes, a global $U(1)$ symmetry called
Peccei-Quinn symmetry is presumably needed to explain why
the strong interaction is CP invariant. To avoid conflict with
accelerator physics and astrophysics, the symmetry is broken at
a very high energy scale $\gsim 10^{10}\GeV$.
\item
Going to a still higher scale, one would like to make contact with
gravity, presumably involving superstring degrees of freedom
around the Planck scale $\sim 10^{19}\GeV$.
\item
One may wish to invoke a GUT. Superstring theory can apparently
relate couplings etc without one, but on the other hand
the ratio of the
strong, weak and electromagnetic interaction
strengths predicted by supersymmetric GUTS agrees with experiment
amazingly well, with a unification scale $\simeq 10^{16}\GeV$.
\end{itemize}
An observational reason for wanting to extend the Standard Model
might be the desire to introduce neutrino masses to solve the solar
neutrino problem as well as a problem with atmospheric neutrinos.

Because the cosmological predictions of the Standard Model agree so well
with observation, many otherwise reasonable
extensions of the Standard Model can be ruled out.
Particularly sensitive are nucleosynthesis, and the absence of
significant corrections to the shape of the cmb spectrum.

{}From the many things that could be discussed, I choose three of the
most topical.

\subsubsection*{Neutrino masses}

A massive neutrino species could be a dark matter candidate.
{}From \eqss{11}{31}{48a},  its contribution to the density parameter is
\be
\Omega_\nu=.109h\mtwo \rfrac{m_\nu}{10\eV}
\label{65}\ee
A mass of order $10\eV$ is therefore required. Experiments observing
$\beta$-decay
practically rule out such a mass for the $\nu_e$, but it
is allowed for the $\nu_\mu$ or $\nu_\tau$.

A firmer reason for wanting some neutrino species to have mass comes
from the solar neutrino problem, whose favoured solution requires
either the $\nu_\mu$ or the $\nu_\tau$
to have a mass of about $3\times 10^{-3}\eV$ (with the $\nu_e$ mass much
smaller).

A natural mechanism for generating neutrino masses is the see-saw
mechanism, which invokes a right-handed neutrino with very large
mass $M$, related to the lepton mass $m$ and neutrino mass $m_\nu$
of each generation by
\be
m_\nu=m^2/M
\label{66}\ee
Taking, without very good justification, $M$ to be the same for all
three generations, the choice $M\sim 10^{12}\GeV$ gives
$m_{\nu_\tau}\sim 10\eV$ and $m_{\nu_\mu}\sim 3\times10^{-3}\eV$,
so that the first neutrino can be the dark matter and the second can
solve the solar neutrino problem.

Neutrino dark matter is called hot, because it remains relativistic
until the epoch when most cosmological scales of interest have entered
the horizon (come into causal contact), and therefore cannot initially
undergo gravitational collapse.

\subsubsection*{Cold and warm dark matter}

Cold dark matter is by definition non-relativistic when all cosmological
interesting scales enter the horizon. The lightest supersymmetric
particle is a cold dark matter candidate. Its interaction strength is
similar to that of the neutrino, but because it is much
heavier it will be non-relativistic when it falls out of equilibrium.
Its contribution to $\Omega_0$ turns out to be
\be
\Omega\sub{lsp}\sim \rfrac{m}{1\TeV}^2 h\mtwo
\label{67}\ee
Since one needs $m\sim100\GeV$ to $1\TeV$ for supersymmetry to work,
this is automatically of order 1!

Warm dark matter by definition remains relativistic until a
cosmologically interesting epoch, which is however significantly earlier
than the epoch for neutrino dark matter. Candidates include a right
handed neutrino, a majoran, or a particle with interaction strength very
much less than that of neutrinos with a mass of order $1\keV$.

The dark matter candidates discussed so far are known collectively as
WIMPS (weakly interacting massive particles). One can hope to detect
a WIMP in the laboratory by searching for its rare interactions, the
detectors being of the same type as the ones that see solar neutrinos
or do neutrino astronomy (so far only with supernova 1987A), and place
limits on proton decay. One might also hope to detect WIMPS in the
galaxy by looking for photons produced by their annihilation. Finally
one might be able to create them at accelerators.

A quite different dark matter candidate is the axion, which is never in
thermal equilibrium and would be cold dark matter.
 The axion is the Goldstone boson of Peccei-Quinn
symmetry, which is the best bet for ensuring the CP invariance of the
strong interaction (except perhaps for making one of the quark masses
zero). Various possibilities exist for axion cosmology in different
regimes of parameter space \cite{axionme}, but typically
axions are emitted by Peccei-Quinn strings at $T\sim1\GeV$ and quickly become
non-relativistic. The axion
contribution to
$\Omega_0$ is roughly
\be
\Omega_a\sim (10^{-4}\eV)/m_a
\label{71a}
\ee
Accelerator
experiments plus the requirement that axion emission should not have
drained too much energy from supernova 1987A require that $m_a
<10^{-3}\eV$. Since $\Omega_a\lsim 1$,
we conclude that if the axion exists at all its mass is roughly of order
$10^{-3}$ to $10^{-4}\eV$, and it gives $\Omega_a$ very roughly of order
1! One might detect solar axions through their tiny electromagnetic
interaction, by exploiting the fact that their wavelength
$(10^{-4}\eV/m_a)\times .197\cm$ is macroscopic.

In the context of supersymmetry the axion must have two supersymmetric
partners, the axino (spin $1/2$) and the saxino (spin $0$),
with rather well defined properties. The
former might be a dark matter candidate and the latter
might dilute the entropy by its decay, reducing
predicted cosmological mass densities like \eqs{71}{71a}
by a factor of up to $10\mfour$ \cite{saxinome}.

These dark matter candidates are relatively uncontroversial, involving
rather minimal extensions of the Standard Model. More ambitious
extensions can produce dark matter with practically any desired
properties; thus, since MDM (mixed, {\it ie} hot plus cold,
 dark matter)
has proved so successful at fitting the observations, no fewer than
three separate proposals have been made for naturally generating
it in the early universe \cite{silvio2,chun,kaetal}.

A completely different possibility is that the non-baryonic dark matter
consists of black holes which form before nucleosynthesis.
In that case one has no need of non-baryonic dark matter particles.
A black hole will form if there is an overdense region with density contrast
$\delta\rho/\rho\sim1$, whose size is of order the Hubble distance.
All of the energy in the
overdense regions then ends up in the black hole, whose
mass is therefore
equal to $\rho_r/\rho_m$ times the
mass of the matter in a Hubble volume,
\be
M\sim 10\mone (\rho_r/\rho_m) \rfrac{1\MeV}{T}^3
\sim 10^5\rfrac{1\MeV}{T}^2
\label{68}\ee
As we shall see later, the density contrast
at horizon entry is only of order $10^{-5}$ on the scales
$M\gsim 10^{10}\msun$ explored by large scale structure and the
cmb anisotropy, and simple inflation models predict that this remains
true down to very small scales. The simple models could be wrong
however. One might also form smaller black holes
on scales much less than the horizon size at a phase transition.
Light enough black holes could be evaporating during nucleosynthesis,
or at the present, with dramatic effects.

Finally, I note that although WIMP or axionic dark matter is normally
assumed to be gaseous, it might be bound into macroscopic objects.
To achieve this during radiation domination one needs
an isocurvature matter
density perturbation (one not shared by the radiation) with most of its
power on small scales. Such a perturbation might be generated during
inflation, like the usually considered adiabatic density perturbation,
but it could also be generated at a phase transition by, say, a
collapsing bubble.
In that case the maximum mass of the objects formed is that
within a Hubble volume,
\be
M\sim .1 (1\MeV/T)^3 \msun
\label{69}\ee
This mechanism has been considered in the context of axions,
where the phase transition is at $T\sim1\GeV$ giving objects
with $M\sim10^{-10}\msun$ \cite{rocky}.

\subsubsection*{Topological defects}

In the early universe, the Higgs fields (and any other scalar fields
present) can find themselves in a false vacuum, corresponding to a
local minimum of their potential.
The false vacuum is typically stabilized by finite temperature effects,
but an interaction with the inflaton field can also to the job \cite{
LIN2SC,LL2,LIN2SC2,latest,MML}.
When the temperature, or the inflaton field, falls below some critical
value, the false vacuum is destabilized and there is a phase transition
to the true vacuum. Within the Standard Model the electroweak transition is
the only example of this phenomenon, but further transitions could
occur, at much higher energy scales, in extensions of it.

Such a phase transition can leave behind it topological defects,
consisting of regions of space where the field is still trapped in the
false vacuum. Whether this happens or not depends on the nature of the
symmetry breaking. No topological defects form at the electroweak
transition, but they might well form in transitions occurring in
extensions of the Standard Model. A concrete example is provided by
Peccei-Quinn symmetry breaking, which produces global strings which play
an important role in producing axions before they disappear
at $T\sim1\GeV$. Another example is the breaking of a GUT which
inevitably produces magnetic monopoles. They are so abundant as to be
cosmologically forbidden unless one either finds a way of
annihilating them, or invokes inflation.

\section{The Evolution of the Density Perturbation}

Cosmological perturbation theory develops linear equations for
perturbations away from homogeneity and isotropy. Using it one can
follow their growth on a given scale, until they become
big enough for gravitational collapse. On scales $\gsim100\Mpc$, where
collapse has yet to occur, cosmological perturbation theory can be used
right up to the present epoch. On smaller scales the only sure-fire
way of performing calculations after perturbation theory fails is to
perform numerical simulations, though analytic approximations
can provide some insight. In these lectures I concentrate on
the  linear regime.

In the Newtonian regime cosmological perturbation theory has long been
recognised to be a straightforward application of fluid flow equations,
taking into account where necessary particle diffusion and
free-streaming (collisionless particle movement). In the relativistic regime,
cosmological perturbations was first discussed by Lifshitz
 \cite{lifs}
in 1946. His formalism, which has been the starting point for  most
subsequent work including the influential `gauge invariant' formalism of
Bardeen \cite{bardeen,kosa}, considers the perturbed
Robertson-Walker metric. An alternative formalism, which
makes no mention of the metric perturbation and works instead with
relativistic fluid flow equations, was initiated by Hawking
\cite{hawk} in 1966.
This approach, which treats the Newtonian and relativistic regimes
in a unified way, is becoming increasingly popular
\cite{lymu,elbr,lyst,bret,duet}
and is the one
that I will use in these lectures.

\subsection{Relativistic fluid flow}

We populate the universe with comoving observers, who define physical
quantities in their own region. By definition, the momentum
density is zero with respect to a
comoving observer; such an observer moves with the energy flow.

A crucial concept is the {\it velocity gradient} $u_{ij}$. It is
defined by each comoving
observer, using locally inertial coordinates in which he is
instantaneously at rest, as the gradient of
the velocity $u^i$ of nearby comoving worldlines,
\be
u_{ij}\equiv \partial_j u^i
\label{70}\ee
The velocity gradient can be uniquely decomposed into an antisymmetric
vorticity $\omega_{ij}$, a symmetric traceless shear $\sigma_{ij}$,
and a locally defined Hubble parameter $H$,
 \be
u_{ij}=H \delta_{ij}+\sigma_{ij} +\omega_{ij}\;. \label{71}
 \ee
In the limit of homogeneity and isotropy,
$\sigma_{ij}=\omega_{ij}=0$.
One can show from angular momentum conservation that $\omega_{ij}$
decays like $(\rho+p)a^{-5}$, so it is presumably negligible.
We will see how to calculate $\sigma_{ij}$ in Section 3.3.

Just as for a homogeneous isotropic universe,
it is useful to consider
`comoving hypersurfaces', defined as those orthogonal the fluid flow
worldlines.\footnote
{As noted later this definition has to be modified if
$\omega_{ij}$ does not vanish.}
On a given hypersurface, each
quantity
$\rho$, $p$ and $H$ can be split into
an average plus a perturbation,
\bea \rho(\bfx,t)&=& \bar\rho(t)+\delta\rho(\bfx,t)
\label{110} \\
p(\bfx,t)&=& \bar p(t)+\delta p(\bfx,t)
\label{110a} \\
H(\bfx,t)&=& \bar H(t)+\delta H(\bfx,t)
\label{110b} \eea
Here $t$ is the time coordinate labelling the hypersurfaces, and ${\bf
x}=(x^1,x^2,x^3)$ are space coordinates. We would like to choose the
space coordinates to be
comoving coordinates, related to Cartesian coordinates by $r^i=a x^i$, with
$a$ the average scale factor given by $\dot a/a=\bar
H$ (we shall not have occasion
to define a perturbed scale factor). This cannot be done exactly, because
the expansion is not isotropic and the
comoving hypersurfaces are not flat.  However
the departure from flatness is of first order, and can therefore be ignored
when describing perturbations which are themselves of first order. In other
words {\it all perturbations `live' in flat space.}

\subsubsection*{Independent scales}

Each perturbation $f$ can be written as a Fourier series,
defined in a comoving box much bigger than the observable
universe
\be f({\bfx},t)=\sum\sk f\sk(t)
	e^{i
{\mbox{\scriptsize{\bf k}.{\bf x}}}
} \label{116} \ee
The beauty of this expansion is that each Fourier mode propagates
independently, as long as the cosmological perturbation theory that we are
developing here is valid. The inverse wavenumber $a/k$ is said to define a
scale, which is specified by giving its present value $k\mone$.

Now consider a small density enhancement in the early universe,
which is destined to become,
say, a galaxy. If its size is
of order $r=xa$, it is typically made out of Fourier components
with $k\sim x\mone$. As long as it corresponds to a small
density contrast, $\delta\rho/\rho\ll  1$,
it will expand with the universe so that its comoving size
$x$ remains constant. When its density contrast becomes of order 1
it will collapse and then its {\it physical} size will remain more or
less constant. In both cases, though, the {\it mass} of the enhancement
remains fixed. It is therefore useful to associate with each scale
$k$ the mass of matter enclosed within a sphere of comoving radius
$x=k\mone$ (taking the universe to be practically homogeneous,
corresponding to the early universe). This mass is
\be
M(x) =1.16\times 10^{12} h^2 (x/1\Mpc)^3 \msun
\label{76}\ee
One expects perturbations with comoving wavenumber
$k$ to be relevant for the formation of structures with mass $M(x)$,
where $x=k\mone$.

\subsubsection*{Horizon entry}

The ratio of a given comoving scale
$a/k$ to the Hubble distance $H\mone$ is equal to
$aH/k=\dot a/k$, which decreases with time. At the epoch when
this ratio falls
through 1, the scale is said to {\it enter the horizon}.

Well after horizon entry, the scale is small compared with the Hubble
distance, which means that ordinary physical effects like diffusion,
free-streaming (collisionless particle movement)
and the propagation of sound waves can operate, with the expansion of
the universe playing only a minor role. Well before horizon entry,
the scale is much bigger than the Hubble distance, which means that
causal processes of this kind cannot
operate  (at least during
a Hubble time, and in practice not at all).
Instead, as we shall see, each part of the universe evolves
independently.

The scale entering the horizon at a given epoch is given by
\be
k\mone=(aH)\mone=\frac{a_0 H_0}{aH}H_0\mone
\label{77}\ee
Except around matter-radiation equality
at $z\sim 10^4$ one has
\bea
aH&\propto a\mone\hspace{10mm}&\mbox{radiation domination}
\label{78}\\
aH&\propto a\mhalf\hspace{10mm}&\mbox{matter domination}
\label{79}\eea
Thus a crude estimate is that the scale entering the horizon at
$z\lsim 10^4$ is $k\mone\sim z\mhalf H_0\mone$, making the scale
entering the horizon at matter-radiation equality
$k\sub{eq}\mone\sim 10\mtwo H_0\mone$, and
and that the scale entering the horizon
at $z\gsim 10^4$ is $k\mone(z)\sim10^2 z\mone H_0\mone$.
An accurate calculation shows that $k\sub{eq}\mone=40h\mone\Mpc $,
and that the scale entering the horizon at photon
decoupling is $k\sub{dec}\mone=90h\mone\Mpc$.
We shall see that the first scale is crucial for structure formation,
and the second for the cmb anisotropy.
The smallest scale directly relevant for structure formation is presumably
the one corresponding to a dwarf galaxy, which has mass
$M\sim 10^6\msun$ and from \eq{76} corresponds to $k\mone\sim
.01 \Mpc$, which enters the horizon when $T\sim 10\keV$.

\subsubsection*{The evolution of the density perturbation}

Now I derive differential equations for the perturbations. In doing so
one has to remember that the comoving worldlines are not in general
geodesics, because of the pressure gradient. As a result,
the proper time interval $d\tau$ between a pair of
comoving hypersurfaces is position dependent. Its average may be identified
with the coordinate time interval $dt$,
 and one can show (using essentially the Lorentz transformation
between nearby observers) that
its variation with position is given by \cite{lyst,LL2}
\be \frac{d\tau}{dt}=\left(1-\frac{\delta p}{\rho+p} \right)
\label{111} \ee

Along each worldline the rate of change of $\rho$ with respect to proper
time $\tau$ is given by energy conservation and has the same
form \eq{50} as in the unperturbed case,
\be
\frac{\diff \rho}{\diff \tau}
=-3H(\rho+p)
\label{81}\ee
The rate of change of $H$ is given by the Einstein field equation,
and to first order receives just one extra term in the presence of
perturbations, coming from the pressure gradient \cite{lyst},
\be
\frac{\diff H}{\diff \tau}=
-H^2-\frac{4\pi G}{3} (\rho+3p) -\frac13 \frac
	{\nabla^2 \delta p}{\rho +p} \label{112}
\ee
This equation is called the {\it Raychaudhuri equation}.
The operator $\nabla^2$ is the Laplacian on a comoving hypersurface,
given in terms of
comoving coordinates by
\be \nabla^2=a\mtwo \delta^{ij} \pdif{}{x^i}\pdif{}{x^j}
\label{113} \ee

Perturbing $H$, $\rho$ and $p$ to first order and using
\eq{111} gives the following equations for the Fourier
components
\bea (\delta \rho\sk)\dot{}&=& -3(\rho+p)\delta H\sk-3H \delta\rho\sk
\label{114} \\
(\delta H \sk)\dot{}&=& -2 H \delta H\sk
-\frac{4\pi G}{3}\delta\rho\sk
+\frac13\rfrac{k}{a}^2\frac{\delta p\sk}{\rho+p}
\label{117}
\eea
Eliminating $\delta H\sk$ with \eq{114} gives
a second order differential equation for $\rho\sk$. It is
convenient to use the {\em density contrast}
$\delta\equiv \frac{\delta\rho}{\rho}$, in terms of which the equation
is
\be H\mtwo\ddot{\delta}\sk  +[2-3(2w-c_s^2)]H\mone \dot{\delta}\sk
	-\frac32(1-6c_s^2+8w-3w^2)\delta\sk =
	-\left(\frac{k}{aH}\right)^2 \frac{\delta p\sk }{\rho}
	\label{118} \ee
I have used the notation $w=p/\rho$ and $c_s^2=\dot p/\dot \rho$.

In these equations for the perturbations the distinction between $\rho$
and $\bar\rho$ is not meaningful (similarly for $p$ and $H$),
because we are working only to first
order. For the same reason the distinction between $\tau$ and $t$ is
not meaningful. In order to have linear equations one will therefore
obviously take $\rho$ to mean $\bar\rho$ and and similarly for
$p$ and $H$, and will take the dot to mean differentiation with respect
to $t$. Note that $c_s^2=\dot p/\dot\rho$ is the speed of sound,
because $p$ and $\rho$
vary adiabatically in a homogeneous isotropic universe (no heat flow).

\subsubsection*{The case of zero pressure gradient}

The right hand side of \eq{118}, which involves the pressure gradient,
is negligible after matter domination because the pressure is
 negligible.\footnote
{Except for the baryons on scales below the Jeans scale, and we are
presently assuming
that the dark matter is cold, otherwise it is
modified by free-streaming. Both of these points will be addressed
shortly.}
It is also negligible well before
horizon entry even during radiation domination, because
the gradient is so small.\footnote
{Provided that $p/\rho$ is not extremely large, which is ensured by the
adiabatic initial condition defined shortly.}
When it is negligible, \eq{118} can be reduced to a first order
equation, which has a very simple interpretation
\cite{lyth84,lymu}

The solution and its interpretation hinge on the introduction of
a quantity $K$, defined locally through the Friedmann
equation \eq{7}. Just as in the homogeneous, isotropic case, general
relativity shows that $K/a^2$ is a measure of the curvature of
the comoving hypersurfaces.
In the homogeneous, isotropic case $K$ is time independent
by virtue of the energy conservation \eq{50} and the gravitational
deceleration equation \eq{51}. We have seen that in general the latter is
modified to become the Raychaudhuri equation \eq{112},
 and as a result $K$ is not
in general time independent.
But when the pressure gradient is negligible they are both
unmodified, so that $K$ is time independent. We can say that
{\it when the pressure gradient is negligible, each region of
space evolves like a separate Friedmann universe.}

On a comoving hypersurface $K$ can be split into
and average $\bar K$ plus a perturbation $\delta K$, but the average can
be set equal to zero because $\Omega\simeq1$.
Perturbing the Friedmann equation therefore gives, to first order,
\be 2H\delta H\sk =\frac{8\pi G}{3} \delta \rho \sk
-\frac {\delta K\sk }{a^2} \label{120}  \ee

When $\delta K\sk$ is time independent,
\eqs{114}{120} give a {\it first} order differential equation
for the density contrast,
\be \frac{2H\mone}{5+3w} \frac{d}{dt}
\left[ \rfrac{aH}{k}^2 \delta\sk \right]
+ \rfrac{aH}{k}^2 \delta\sk
=\frac{2+2w}{5+3w} {{\cal R}}\sk  \label{121} \ee
where $w=p/\rho$ and I have introduced the useful quantity
\be {{\cal R}} \sk =\frac32 \frac{\delta K\sk }{k^2}
\label{90}\ee

Remembering that $\delta K/a^2$ is the curvature perturbation and that
it has units of (length)${}\mtwo$, we see that $\cal R\sk=(3/2)
(\delta K/a^2)
(a^2/k^2)$ essentially measures the curvature perturbation in units
of the relevant scale $a/k$. (The factor $3/2$ is chosen so as to give
the simple relation \eq{calr} with the inflaton field perturbation.)
As we shall verify shortly, another interpretation of $\calr$ is that it
is essentially the Newtonian gravitational potential caused by $\delta
\rho$.

During any era when $w$ is constant, \eq{121} has the solution
(dropping a decaying mode)
\be \rfrac{aH}{k}^2 \delta\sk =\frac{2+2w}{5+3w} {{\cal R}} \sk
\label{123} \ee
In the radiation dominated era before horizon entry this becomes
\be
\rfrac{aH}{k}^2 \delta\sk =\frac49 {{\cal R}}\sk (\mbox{initial})
\label{125} \ee
and in the matter dominated era it becomes
\be \rfrac{aH}{k}^2 \delta\sk = \frac25 {{\cal R}}\sk (\mbox{final})
\label{126} \ee
As the labels imply I am regarding the value of $\calr\sk$
during the first era as an `initial condition', which determines
its value during the `final' matter dominated era.

For future reference note that during matter domination,
$H\propto t\mone\propto a^{-3/2}$ and
\be
\delta\sk\propto a \hspace*{1cm}\mbox{(matter domination)}
\label{126a}\ee

\subsection{The transfer function}

For scales entering the horizon well after matter domination
($k\mone\gg k\mone\sub{eq} =40h\mone
\Mpc$) the initial and final eras overlap so that
${{\cal R}}\sk (\mbox{initial})={{\cal R}}\sk (\mbox{final})$.
Otherwise there is a {\it transfer function} $T(k)$,
which may be  defined by
\be {{\cal R}}\sk (\mbox{final})= T(k) {{\cal R}}\sk (\mbox{initial})
\label{127} \ee
An equivalent, and more usual, definition is
\be a\mone \delta\sk \mbox{(final)}
	=A T(k) \delta\sk \mbox{(initial)}
 \label{127a} \ee
where the (time dependent) right hand side is evaluated at an
arbitrarily chosen time during the initial era, and the constant $A$ is
chosen so that $T$ becomes equal to 1 on large scales.

\subsubsection*{The adiabatic initial condition}

In order to calculate the transfer function one needs an initial
condition, specifying the density contrast of each species of matter and
radiation before horizon entry and before matter domination.
The matter consists of baryons and one or more species of non-baryonic
dark matter.
The radiation consists of photons and massless neutrinos.

The most natural condition, is the
{\em adiabatic} condition, which is that the density of each species
depends only on the total energy density, or equivalently on the temperature.
In other words, each density contrast vanishes
on a hypersurface of constant energy
density. Going along any comoving worldline, the density $\rho_r$ of any
species of radiation varies like $a\mfour$ and the density $\rho_m$ of
any species of matter varies like $a\mthree$, so that
\be
\delta\rho_m/\rho_m= \frac34\delta\rho_r/\rho_r
\label{95}\ee
Thus, the adiabatic initial condition implies that each species of
radiation has a common density contrast and so has each species of
radiation, and that they are related by a factor $3/4$.

The most general initial condition is a superposition of the adiabatic
one and some {\it isocurvature} initial condition, specifying a set of
initial density perturbations $\delta\rho_i$
which add up to zero. An isocurvature
initial condition
is not very natural, especially in the context of inflation.
Furthermore, it turns out that a purely isocurvature intial condition
(with a reasonably flat, Gaussian spectrum) cannot
lead to a viable structure formation
scenario.
The adiabatic condition is assumed from now on.

\subsubsection*{Cold and hot dark matter}

The other ingredient needed to calculate the transfer function
is the nature of the dark matter.
It is practically always taken to have negligible interaction with
anything else (to keep it dark). Neutrino dark matter is called
hot dark matter (HDM) because it remains relativistic
(in other words it is radiation rather than matter) until a very later
epoch, which is after structure forming scales have entered the horizon
for pure HDM. Cold dark matter (CDM) by contrast is nonrelativistic at all
epochs of interest, and warm dark matter is in between.
Only CDM and HDM
have as yet been studied in the light of modern data. Pure HDM
is ruled out (at least if we do not form structure with seeds such
as cosmic strings, or with a density perturbation having
so much small scale power that very small objects form first).
Pure CDM
performs rather impressively, going a long way towards explaining
several different types of data, but it cannot actually fit them all
simultaneously within their accepted uncertainties. For this
reason people have considered {\it mixed dark matter} (MDM),
roughly 70\% cold and 30\% hot, which at the time of writing seems to do
rather well \cite{mdm,LL1,bobqaisar}.

\subsubsection*{The CDM transfer function}

The transfer functions for pure CDM and for MDM
are shown in Figure 2. Although several different effects have to be
taken into account to calculate them accurately, their  general form
is easy to describe. Let us deal first with the pure CDM transfer
function.

\begin{figure}
\vspace*{10cm}
\caption[The transfer function]
{The transfer function $T(k)$, taken from  Pogosyan and Starobinsky
\cite{mdm}. From top down, the curves correspond to $\Omega_\nu
=0,0.05,...0.95$. The transfer function is evaluated at the present
epoch, but in the limit of pure CDM (top curve) it becomes time
independent.}
\end{figure}

On a given scale essentially nothing happens before
horizon entry because there is no time for particles to move
on that scale. As a result, the growth
$\delta\propto(aH)\mtwo$ of the density contrast, that we established
earlier as a consequence of the `separate Friedmann universe' evolution
of each comoving region of space, is shared by each species.
After horizon entry, the density contrast of
massless neutrinos rapidly {\it decays away}, because
the neutrinos {\it free-stream}
(that is, travel freely) out of any over dense region. The baryons and photons
are thermally coupled until the epoch of photon decoupling,
and their density contrast
{\it oscillates} as a sound wave, which decays more or less rapidly
(depending on the scale)
as the photons diffuse away from overdense regions.
The density
of cold dark matter {\it grows}, but until matter domination the amount
of growth is rather small because the matter does not have much
self-gravity. For
example, taking the radiation to be uniform
and ignoring the baryons, one finds by following the steps leading to
{}~\eq{118} that the matter density contrast satisfies
\be \ddot \delta_m + 2 H \dot \delta_m
-4\pi G \rho_m \delta_m =0 \label{133}
\label{96}\ee
Using $\rho_m/\rho_r=a/a\sub{eq}$, one
finds that the solution of this equation is
proportional to $2+3a/a\sub{eq}$ (plus a decaying mode), which indeed hardly
grows before
matter domination at $a=a\sub{eq}$.
Since it would have grown like $(aH)\mtwo$
had horizon entry not occurred, the total amount of growth missed by a
given scale is roughly
\be
\frac{(aH)^2\sub{hor}}{(aH)^2\sub{eq}}=\rfrac{k}{k\sub{eq}}^2
\label{97}\ee
The CDM
transfer function is therefore roughly
\bea
T(k)&\simeq 1 \hspace{2cm} & (k\mone>k\sub{eq}\mone=40h\mone\Mpc)
\label{98}\\
T(k)&\simeq  (k\sub{eq}/k)^2 & (k\mone<k\sub{eq}\mone).
\label{99}\eea

This indeed gives an extremely crude representation of the
correct CDM transfer function.
To calculate it accurately one needs to follow in
detail the oscillation and/or decay of each component.\footnote
{A surprisingly good approximation \cite{machacek}
is to ignore the decay, treating
the radiation and matter as uncoupled perfect fluids
\cite{lyst,duet}, but I do not
know of any simple reason why this is so.}

The above discussion applies only to the CDM density perturbation.
 We need to know also what happens to the baryon density perturbation,
since baryons after all are what we observe.
Unlike the CDM
density contrast,
the baryon density contrast is small at the photon decoupling epoch
because it has been decaying since horizon entry.
Afterwards, the baryons are unaffected by the photons, but
two competing forces act on them.
First there is gravity, which tends to make them fall into the
potential wells caused by the CDM density contrast, and second there is
their own pressure gradient which tends to keep them out of the wells.

To rigorously see which effect wins one should generalize \eq{118} to
treat the CDM and the baryons as a pair of
uncoupled fluids \cite{lyst,duet}.
 In practice the following order of magnitude estimate
is enough.
Ignoring the pressure, the time taken for the baryons to fall
into a well is\footnote
{If a particle falls from rest towards a point mass
$M$, its velocity at distance $r$ is given by $mv^2=2GM/r$
so it falls a significant distance in a time $t
\sim r/v\sim (GM/r^3)\mhalf$. We are replacing the point mass
by a perturbation with size $r$ and density $\rho\sim M/r^3$.}
 of order $(G\rho)\mhalf$.
The time taken for the pressure to adjust itself to prevent the collapse
is of order $\lambda/c_s$ where $\lambda=2\pi/k$ is the wavelength and
$c_s$ is the speed of sound.
Collapse occurs if $\lambda/c_s\lsim(G\rho)\mhalf$
because the pressure cannot act quickly enough.
Inserting by convention a factor 2, one
concludes that collapse occurs on scales
in excess of
$k\sub{J}=(4\pi G \rho/v_s^2)\half$.
This is called the {\it Jeans scale} and the
corresponding mass, given by \eq{76}, is called the Jeans mass.

Just after decoupling the speed of sound is given by
$c_s^2=5T/3m_N$ where $T\simeq 1100\times 2.74$\,K. The
corresponding Jeans mass is of order $10^6\msun$.
As yet structure formation on such small scales is not well understood
because the necessary numerical simulations of the collapse are too
difficult, but it may be significant that there are no galaxies with
mass less than the Jeans mass.

\subsubsection*{The MDM transfer function}

The hot (neutrino) component
becomes non-relativistic only at the epoch
\be 1+z\sub{nr}=(1.7\times 10^5)h^2\Omega_\nu \label{200}\ee
On scales  entering the horizon before this epoch, the neutrino
dark matter will free-stream out of any density enhancement
so that its density contrast will decay. Afterwards
the HDM density contrast is free to
grow to match that of the CDM, on scales in excess of an effective
Jeans length
\be k_J\mone=.11(1+z)\half\Omega_\nu \Mpc \label{202}\ee
By the present time, the CDM  and HDM have a common density contrast on scales
$k\mone\gsim .1\Mpc$, which are the only ones of interest. Note, though,
that the MDM continues to evolve right up to the
present.

\subsection{The peculiar velocity field}

After matter domination it is very useful to introduce the concept of a
peculiar velocity field. Traditionally this concept is formulated in
the context of the Newtonian treatment of perturbations, but it has been
recognised recently that the same concept is valid, and useful, also in the
general relativistic treatment \cite{LL2,brly}.

At a given epoch the
Newtonian treatment is valid on scales much less than the Hubble
distance, to the extent that gravity waves can be ignored. There
is a well defined fluid velocity $\bfu$, and choosing the reference
frame so that $\bfu$ vanishes at the origin the
PV field
${\mbox{\bf v}}$ is defined\footnote
{Up to a constant, which can be
chosen so that the average of ${\mbox{\bf v}}$ vanishes. }
 by
\be
{\mbox{\bf u}}({\mbox{\bf r}})-{\mbox{\bf u}}(0)= \bar H {\mbox{\bf r}}+
{\mbox{\bf v}}(
{\mbox{\bf r}})-{\mbox{\bf v}}(0)\;,
\label{203}
\ee
An equivalent statement in terms of the velocity gradient \eq{71} is
\be
\delta u_{ij}=\partial_i v_j\;.
\label{9a}
\ee
where $\partial_i=a\mone\partial/\partial x^i$.
Like any vector field, $\bfv$ can be written
${\mbox{\bf v}}={\mbox{\bf v}}\su{L}+{\mbox{\bf v}}\su{T}$, where
the transverse part ${\mbox{\bf v}}\su{T}$
satisfies $\partial_i v\su{T}_i=0$
and the longitudinal part is of the form
 ${\mbox{\bf v}}\su{L}={\mbox{\boldmath $\nabla$}}
\psi_v$. \eq{14} defines the local Hubble parameter $H$, shear
$\sigma_{ij}$ and vorticity $\omega_{ij}$, this last being given by
\be
\omega_{ij} =
\frac12 (\partial_i v\su{T}_j-\partial_j v\su{T}_i)
\label{203a}\ee
Angular momentum conservation gives
$\omega_{ij}\to a\mtwo$, so $\bfv\su{T}$ decays like $a\mone$
and may be dropped (remember that $\partial_i\equiv
a\mone\partial/\partial x^i$).

Taking $\bfv$ to be purely longitudinal, it is determined
by the density perturbation in the following way.
First, take the trace of \eq{14} to learn that
$\del. {\bfv} = 3\delta H$. From
\eqsss{15a}{120}{90}{123}, it follows that
\be \del. {\bfv}=-(4\pi G\delta\rho ) t \label{147}\ee
The solution of this equation is
\be {\bfv}=-t\del\psi
\label{147a} \ee
or
\be v_i(\bfx,t)=- (t/a)\pdif{\psi(\bfx) }{x^i} \label{147b}\ee
where
\be \psi(\bfx) = - G a\mtwo \int\frac{\delta\rho(\bfx^\prime ,t)}
	{|\bfx^\prime -\bfx|} d^3x^\prime  \label{148} \ee
The factor $a\mtwo$ converts coordinate distances into physical distances.
Since it is related to the density perturbation by the Newtonian expression,
$\psi$ is called the peculiar gravitational potential. It is independent of
$t$ because, from \eq{126a},  $\delta \rho\propto a^2$.

{}From \eq{126} we see that the peculiar gravitational potential is
related to the spatial curvature perturbation by
\be \psi= - \frac35 {{\cal R}}(\mbox{final})  \label{149} \ee

{}From \eqs{147a}{148} the Fourier components of $\bfv$, $\psi$ and $\delta$
are related by
\bea {\bfv}\sk &=& i\frac{\bfk}{k}\rfrac{aH}{k} \delta\sk
\label{159}\\
\psi\sk &=& -\frac32 \rfrac{aH}{k}^2 \delta\sk
\label{159a}\eea

The extension of these results to the relativistic regime turns out to
be very simple \cite{LL2,brly}.
 The vorticity decays $(\rho+p)a^{-5}$ and is again
presumably negligible.
In that case one can show that the velocity gradient perturbation
is of the form
\be
\delta u_{ij} =\partial_i  v_j +
\frac 12 \dot h_{ij}
\label{duij}
\ee
The extra term, which is absent in Newtonian physics, represents the
effect of gravitational waves. The success of Newtonian theory
on scales $\lsim100\Mpc$ implies that it
is negligible there, but
as we shall see it could be significant on larger scales,
and contribute to the cmb anisotropy. Even if present though, it
does not spoil \eqst{147}{159a} because it
 is transverse, $\partial_i h_{ij}=0$, and
traceless, $\delta^{ij}h_{ij}=0$.

One can avoid dropping the
vorticity by proceeding as follows \cite{brly}.
First, a transverse velocity
$\bfv^T$ may be {\it defined} by \eq{203a}.
The worldlines with velocity
$-\bfv^T$ relative to the comoving worldlines
have no vorticity, and comoving hypersurfaces are defined to
be orthogonal to them (there are no hypersurfaces orthogonal to
worldlines with nonzero vorticity).
The velocity gradient $\delta u_{ij}$ receives an extra contribution
$
\frac12 (\partial_i v\su{T}_j - \partial_j v\su{T}_i)+
\frac12 (\partial_i w\su{T}_j + \partial_j w\su{T}_i)
$ where
$
\mbox{\bf w}\su{T}= \left[1+6\left(1+\frac{p}{\rho} \right)
\left(\frac{aH}{k}\right)^2 \right] {\mbox{\bf v}}\su{T}
$
For scales well inside the horizon $\bfw$ is negligible and the
Newtonian result is recovered.

\subsection{The spectrum of the density perturbation}

In order to discuss the perturbations in a given region of the universe
around us, one has to perform the Fourier expansion \eq{116} in a
box much bigger than this region. If the box is a cube with sides of
length $L$, the possible values of $\bfk$
form a cubic lattice in $k$ space with spacing $2\pi/L$.

When discussing an isolated system, which is the usual case in physics,
one can take the limit $L\to\infty$ in a straightforward way, the
coefficients $f\sk$ tending to a constant limit which is a smooth function
of $\bfk$. But cosmological perturbations do not fall off at large distances,
and their Fourier coefficients are not smooth functions of $\bfk$.
They are the spatial
analogue of a signal extending over an indefinite period of time, as
opposed to an isolated pulse.

Although the coefficients $f\sk$ are not smooth functions, it is
reasonable to suppose that $|f\sk|^2$ is smoothly varying when
smeared over a region $\diff^3k$ of $k$ space, which is large enough to
contain many lattice points. I shall denote this average by $\langle
|f\sk|^2\rangle$. It depends only on $k=|\bfk|$, and up
to a $k$ dependent factor it is called the {\it spectrum}
of $f$, because of the analogy with a signal. A convenient choice of the
factor is to define the spectrum as
\be {{\cal P}}_f\equiv \rfrac{Lk}{2\pi}^3 4\pi \langle|f\sk|^2\rangle
	\label{214} \ee
The normalisation is chosen to give a simple formula for the dispersion
(root mean
square) of $f$, which I shall denote by $\sigma_f$. From the Fourier
expansion  one has $\sigma_f^2=\sum |f\sk^2|$, and since the possible
values of $\bfk$ form a cubic lattice with spacing $2\pi/L$ the transition
from sum to integral is
\be \rfrac{2\pi}{L}^3\sum\sk \longrightarrow 4\pi \int k^2  dk
\label{215}\ee
The dispersion $\sigma_f$ is therefore given by
\be
\sigma_f^2 \equiv\langle f^2(\bfx) \rangle
= \int^\infty_0 {{\cal P}}_f(k) \frac{dk}{k}
\label{DISP}
\ee
with the brackets now denoting the average over position $\bfx$.

For the density perturbation $f=\delta$ it is useful to define the
correlation function $\xi(r)$ by
\be \xi(r)=\langle f(\bfr+{\bfx})f(\bfr)\rangle = \int^\infty_0 {{\cal P}}_f(k)
	\frac{\sin(kr)}{kr} \frac{dk}{k}
\label{corr} \ee
The analogous
quantity is useful for
other perturbations like the peculiar velocity components, though it is not
then called the correlation function.
For $r=0$ it clearly reduces to $\sigma_f^2$.

If the phases of the Fourier coefficients are random, $f$ is said to be
Gaussian, and then all of its stochastic properties are determined by its
spectrum. In particular the probability distribution of $f$, evaluated at
randomly chosen points, has a Gaussian profile. As long as the perturbations
are evolving linearly this Gaussian property is true of perturbations
originating as inflationary fluctuations,
and I take it for granted from
now on.

{}From \eqs{126}{127}, the spectrum of the density contrast after matter
domination may be written
\be
{{\cal P}}_\delta(k) =\rfrac{k}{aH}^4 T^2(k) \delta_H^2(k)
\label{23b} \ee
The quantity $\delta_H$ specifies the initial spectrum.
 In fact, from \eq{125}
it is related to the spectrum of the initial curvature perturbation
${{\cal R}}$ by
\be \delta_H^2(k)=\frac{4}{25} {{\cal P}}_{{\cal R}} (k)
\label{24b} \ee
The  subscript $H$ has been used because $\delta_H^2$ is exactly equal to the
value of ${{\cal P}}_\delta$
on horizon entry on scales $k\mone\gg k\mone\sub{eq}$,
and approximately equal to it on smaller scales. As we shall see later,
this means that $\delta_H(k)$ is {\it roughly} equal to the
{\it rms} value of $\delta$ at horizon entry, for a density enhancement with
comoving size of order $k\mone$.

The standard assumption is that $\delta_H^2$ is independent of $k$.
A more general possibility is to consider a spectrum
\be \delta_H^2\propto k^{n-1} \label{220}\ee
where the exponent  $n$ is called the {\it spectral
index}. (The definition of the index as $n-1$ instead of $n$
is a historical accident.)
The standard choice of $n=1$ was first advocated by Harrison (1970)
and Zel'dovich (1970) on the ground that it is the only one making the
perturbation small on all scales, at the epoch of horizon entry.
Although this is a powerful argument in favour of a value of $n$
{\it of order} 1, it does not require that $n$ is actually close to
1 unless one assumes, without justification, that the power law
dependence holds over a huge number of decades of scale\footnote
{On cosmologically interesting scales the
observed magnitude of $\delta_H$ is of order $10^{-5}$, so if
$n=1-\delta n\simeq1$ one has to extrapolate over $5/\delta n$ orders
of magnitude before $\delta_H$ becomes of order 1.}

In this context, inflation makes a striking prediction \cite{LL1}.
It gives the
spectral index is given in terms of two parameters
$\epsilon_1$ and $\eta_1$,
\be
n=1+2\eta_1-6\epsilon_1
\label{221}\ee
The parameters depend on the model of inflation, but both are
much less than 1 in magnitude and $\epsilon_1$ is positive.
As a result most inflationary models predict that $n$ is rather close to
1, and typically smaller rather than bigger. The prediction that $n$ is
close to 1 seems to be confirmed by observation, which is a striking
success for the general idea of inflation. In the very near future,
one will be able to use the observed value of $n$ as a powerful
discriminator between different inflationary models.

\section{The Cosmic Microwave Background Anisotropy}

The first detection last year of the intrinsic anisotropy
of the cmb was surely the
most important advance in observational cosmology for a long
time. Exploring scales two orders of magnitude bigger than
the biggest that can be probed by galaxy surveys, the fact that
it agreed within a factor of two with
extrapolation from these surveys
using the
pure CDM model with a flat spectrum was extremely impressive. Before
exploring the significance of this fact in the next section, let us study
the cmb anisotropy is some detail.

\subsection{The spectrum of the cmb anisotropy}

At a given wavelength, the
cmb is anisotropic {\it ie} its intensity depends on the direction of
observation.
The anisotropy is usually specified by giving the equivalent variation
in the temperature
of the blackbody distribution. Denoting the direction of observation by
a unit vector ${\mbox{\bf e}}$, the anisotropy
may be expanded into multipoles
\be \frac{\Delta T(\bfe)}{T}=\bfw.\bfe+\sum_{l=2}^{\infty}
	\sum_{m=-l}^{+l} a_l^m Y_l^m (\bfe) \label{mult} \ee

The dipole term $\bfw.\bfe$ is well measured, and is the
Doppler shift caused by our velocity $\bfw$ relative to the rest frame
of the cmb (defined as the frame where it has zero momentum density).
Unless otherwise stated, $\Delta T$ will
denote only the non-dipole part from now on. Its
mean square over the whole sky is
\be \left\langle \rfrac{\Delta T}{T}^2 \right\rangle
	=\frac1{4\pi} \sum_{l=2}^\infty\sum_{m=-l}^l |a_m^l|^2
\label{223}\ee

The multipoles, and therefore the mean square anisotropy, depend on
the position of the observer. Averaging over position gives the result
for a randomly placed observer,
\be \left\langle\left\langle \rfrac{\Delta T}{T}^2 \right\rangle
	\right\rangle\sub{position}
	=\frac1{4\pi}\sum_{l=2}^\infty (2l+1) \Sigma_l^2
\label{224}\ee
where
\be \Sigma_l^2=\left\langle |a_l^m|^2 \right\rangle\sub{position}
\label{225}\ee

In contrast with $|a_l^m|^2$, the quantities $\Sigma_l^2$ are
expected to be smooth functions of $l$. For large $l$ one
can
therefore replace the sum by an integral,
\be \left\langle\left\langle \rfrac{\Delta T}{T}^2 \right\rangle
	\right\rangle\sub{position}
	\simeq\frac1{4\pi}\int^\infty_2
l(2l+1)\Sigma_l^2 \frac{dl}{l} \label{4last}
\label{226}\ee
Another nice feature of large $l$ is the relation $\theta \sim 1/l$ between
the angular size of a feature in the sky (in radians) and the order $l$ of the
multipoles that dominate it. (This is  analogous to the relation
$r\sim 1/k$ between the linear size of a feature and the wavenumbers that
dominate its Fourier expansion.) Translating to degrees we have the following
relation between $l$ and the angular scale
\be \frac\theta{1^0}\simeq \frac{60}{l}
\label{227}\ee

By analogy with the \eq{DISP} for the density contrast, one may call
$(2l+1)l\Sigma_l^2/4\pi$ the {\it spectrum} of the cmb anisotropy.

\subsection{Predicting the cmb anisotropy}

The predicted spectrum of the cmb anisotropy is shown in
Figure 3. I will discuss the prediction
qualitatively, then give a
detailed calculation of the large scale (small $l$) part which is the
most interesting.

\begin{figure}
\vspace*{10cm}
\caption[The cmb anisotropy]
{This figure is reproduced from \cite{cretal}.
The top Figure shows the spectrum of
the cmb anisotropy for a slightly tilted standard CDM model with spectral
index $n=0.85$ and $\Omega_B=0.05$. As discussed in the text, the case $n=1$
is recovered for $l\gg1$ by multiplying the curves by a factor
$l^{(1-n)/2}=l^{.075}$. The contribution of an adiabatic density perturbation
is the middle line, labelled `S', and the contribution of gravitational waves
is the bottom line, labelled `T'. The light dashed line is the density
contribution with $\Omega_B=0.01$. For each curve, the quantity plotted is
$l(l+1) \Sigma_l^2$, normalised to 1 for the quadrupole, $l=2$. If the
gravitational wave contribution to the quadrupole is equal to that of the
density perturbation, as is roughly the case for power-law inflation with
$n=0.85$, the top curve indicates the total. On the other hand, it could well
be that the gravitational contribution is negligible. The bottom Figure shows
the filters $F_l$
for various experiments, which are cited in \cite{cretal}.}
\end{figure}

Discounting the possibility of early re-ionisation the cmb last
scattered at the epoch of photon decoupling. Its
surface of last scattering therefore
lies practically at the particle horizon whose comoving distance is
$x=2H_0^{-1}=6000h\mone\Mpc$. At this surface an angle $\theta$ degrees
subtends a comoving distance
\be
\frac{x}{100h\mone\Mpc}\simeq \frac{\theta}{1^0}
\label{228} \ee
This gives, at least roughly, the linear scale explored by
observing the cmb on a given angular scale.

It is convenient (at least pedagogically)
to separate the anisotropy into an initial anisotropy,
specified on a comoving hypersurface shortly after last scattering,
and the additional anisotropy acquired on the journey towards us,
\be
\frac{\Delta T(\bfe)}{T}=\rfrac{\Delta T(\bfe)}{T}\sub{em}+
\rfrac{\Delta T(\bfe)}{T}\sub{jour}
\label{229}\ee
The initial anisotropy represents the variation of the intensity
of the cmb, as measured by different comoving observers, each with their
detector pointing in the relevant direction $\bfe$.

Let us discuss these contributions, from the largest scales downwards.
Scales in excess of a degree or so explore
linear scales
in excess of $100\Mpc$, neatly complementing galaxy surveys which
explore scales $\lsim 100\Mpc$. Such scales have not  entered the
horizon at last scattering, so
the adiabatic initial condition \eq{95}
still holds at that epoch.
 The cmb is practically isotropic for each comoving observer,
 but its
intensity varies from place to place because $\rho_r\propto T^4$.
Since last scattering occurs during matter domination, the
density contrast is practically that of the matter, so the adiabatic
condition gives
\be
\rfrac{\Delta T(\bfe)}{T}\sub{em}\simeq\frac43
\delta(\mbox{\bf x}\sub{em},t\sub{em})
\label{230}\ee
However we shall see that this effect is dominated by the additional
anisotropy acquired on the journey towards us which is called
(on these large angular scales) the Sachs-Wolfe effect.

Scales in the {\it arcminute} regime
explore distance
scales of order 10 to 100 Mpc, the same as the one explored by large scale
galaxy surveys.
As the scale decreases through this regime, the
anisotropy at first exhibits a peak, which can be interpreted as
of the Doppler shift caused by the
peculiar velocity of the initial comoving hypersurface.
Thereafter the anisotropy
acquired on the journey towards us becomes
negligible.\footnote{I here discount the
Sunyaev-Zel'dovich effect by which the cmb acquires anisotropy through photon
up-scattering off dust on its way through a galaxy or cluster. It occurs on
scales of order one arcminute and is easily identified for nearby clusters.
The cumulative effect of distant clusters is expected to be negligible.
} The initial anisotropy also falls, but it
remains significant until the scale falls to a few
arcseconds.

On scales less than a few arcminutes, the distance \eq{228} subtended at the
last scattering surface is less than the thickness of this surface.
As a result
the cmb is isotropised by its passage through the last scattering
surface, and it turns out that no significant
anisotropy is acquired by the cmb on its journey towards us
either. The anisotropy is therefore expected to be very small.

\subsubsection*{The cmb transfer function and the cosmic variance}

By evaluating these physical effects, one can calculate
the multipoles
$a_l^m$ seen by an observer at any given position,
in terms of
the initial density perturbation
$\delta\sk$(initial).\footnote
{Discounting gravitational waves, which I consider in a moment, and
isocurvature density perturbations, cosmic strings etc which I do  not
consider at all.}
Of course we do not know
$\delta\sk$(initial), nor are we interested in it. Instead we know, or
rather are prepared to hypothesise, its
stochastic properties, which are that it is Gaussian with a spectrum
specified by $\delta_H^2(k)\propto k^{n-1}$. The Gaussianity implies that the
real and imaginary part of each multipole
$a_l^m$ has a Gaussian probability distribution
as a function of the observer's position, with no correlation between
them (except for the trivial one $a_l^{m*}=a_l^{-m}$).
The spectrum
$\Sigma_l^2$ of the cmb
anisotropy---specifying the widths of the Gaussian
distributions---is related to the spectrum $\delta_H(k)$ of the
initial density perturbation by a transfer function $T_l(k)$,
\be
\Sigma_l^2=\int^\infty_0 T^2_l(k) \delta_H^2(k) \frac{\diff k}{k}
\label{231}\ee
In accordance with \eq{228}, $T_l(k)$ is peaked at
\be
\frac{l}{60}\simeq \frac{100h\mone\Mpc}{k\mone}
\label{232}\ee
The transfer function is independent of the nature of the dark matter
on large scales as we discuss in a moment. On smaller scales the effect
of adding in some hot dark matter has not yet been investigated in
detail, but going from pure CDM to MDM ($30\%$ hot dark matter)
is not expected to have a big effect. The
prediction in Figure 3 is for pure CDM.

This prediction gives the expectation value $\Sigma_l^2$ of the
multipole
$|a_l^m|^2$ evaluated at a random position, but we can only measure
$|a_l^m|^2$ at our position. The
best guess  is that $|a_l^m|^2=\Sigma_l^2$, but one can also calculate the
expected error, which is the square root of the
variance of the guess (the mean value of
$(|a_l^m|^2-\Sigma_l^2)^2$). This variance is called the
{\it cosmic variance}. For the real or imaginary part of a single
multipole, the Gaussian distribution implies that the variance is
equal to 2 times the mean square prediction. The $l$th multipole has
$2l+1$ independent components, so summing over $m$ makes the cosmic
variance only $2/(2l+1)$ times the mean square prediction.
A typical observation typically involves a (weighted) sum over several
multipoles, which further reduces the cosmic variance so that in
practice it is not usually a dominant factor. One must always bear it in
mind though.

\subsection{The Sachs-Wolfe effect}

Now we calculate the anisotropy on scales in excess of a few degrees
\cite{LL2,brly}.

Consider a photon passing a succession of comoving observers.
 Its trajectory is $a \diff\bfx/\diff t=-{\mbox{\bf e}}$ and
between nearby observers its Doppler shift is
\be
-\frac{d\lambda}{\lambda}=e_i e_j u_{ij} dr=-\frac
{d\bar a}{\bar a} + e_i e_j \delta u_{ij} dr\;,
\ee
where the first term is due to the average expansion, and the second
is due to the relative PV of the observers.
Integrating this expression gives
the redshift of radiation received by us, which was emitted from
a distant comoving source. The unperturbed result is
$\lambda/\lambda\sub{em}=1/a\sub{em}$, and the first order perturbation
gives
\be
\frac{\Delta T(\bfe)}{T}
=\rfrac{\Delta T(\bfe)}{T}\sub{em}+
\int^{x\sub{em}}_0 e_i e_j \delta u_{ij}({\mbox{\bf x}},t) a(t) dx\;,
\label{delt}
\ee
where $x\sub{em}\simeq 2H_0^{-1}$ is the coordinate distance of the
last scattering surface, and the integration is along the
photon trajectory
\be
x(t)=\int^{t_0}_t \frac{dt}{a} = 3
\left( \frac {t_0}{a_0}- \frac ta \right) \label{xtee}\;.
 \ee
Using \eqss{9a}{xtee}{147a} and
integrating by parts one finds
\bea
\frac{\Delta T(\bfe)}{T}&=&
\rfrac{\Delta T(\bfe)}{T}\sub{em}+
\frac 13 [\psi({\mbox{\bf x}}\sub{em} )-\psi( 0)]+
\nonumber\\
&&{\mbox{\bf e}}\cdot[{\mbox{\bf v}}(0,t_0)
-{\mbox{\bf v}}({\mbox{\bf x}}\sub{em},t\sub{em})]\;.
\label{236}
\eea
Here ${\mbox{\bf x}}\sub{em}=x\sub{em} {\mbox{\bf e}}$
is the point of origin of the cmb coming from direction $\bfe$.

A better expression follows \cite{brly} if one uses the divergence theorem and
Eq.~(\ref{xtee}) to project out the dipole part of
$\psi({\mbox{\bf x}}\sub{em})/3$.
One finds that it is equal to
$\langle {\mbox{\bf v}}(t\sub{em})- {\mbox{\bf v}}(t_0) \rangle$
where $\langle\rangle$ denotes the average within the last scattering surface
$x=x\sub{em}$. Defining $\tilde{\mbox{\bf v}}=
{\mbox{\bf v}}-\langle{\mbox{\bf v}}
\rangle$
this gives
\bea
\frac{\Delta T(\bfe)}{T}&=&\rfrac{\Delta T(\bfe)}{T}\sub{em}+
\frac 13 \left[\psi({\mbox{\bf x}}\sub{em} )\right]_{l>1}+
\nonumber\\
&&{\mbox{\bf e}}\cdot\left[ \tilde{\mbox{\bf v}}(0,t_0)-
\tilde{\mbox{\bf v}}({\mbox{\bf x}}\sub{em},t\sub{em})
\right]\;.
\label{best}
\eea

On angular scales $\gg1^0$, corresponding to linear scales
at last scattering which are outside the horizon,
\eqs{159}{159a} show that
\be
|\psi\sk|\gg|\delta\sk|\gg|\bfv\sk|
\label{233}\ee
On these scales we can therefore drop the term $\tilde{\bfv}(\bfx\sub{em}
,t\sub{em})$, as well as the the initial anisotropy \eq{230}.
This leaves
\be
\frac{\Delta T(\bfe)}{T}
=\frac13 \left[\psi({\mbox{\bf x}}\sub{em} )
\right]_{l>1} + {\mbox{\bf e}}\cdot\tilde{\mbox{\bf v}}_0\;.
\label{424}
\ee

The last term in this expression is the dipole and
it defines the rest frame of the cmb by giving our velocity
$\tilde {\bfv}_0=\bfv_0-\langle\bfv_0\rangle$ relative to that frame.
Since $\bfv_0$ is our peculiar velocity, the peculiar velocity of the
cmb rest frame is $\langle \bfv _0\rangle$, the
average peculiar velocity (bulk flow) of the region inside the surface of last
scattering (at roughly the Hubble distance from us).
As we shall discuss in Section 5 the
bulk flow decreases as the scale increases, and is quite negligible
on the Hubble scale, so in practice we can just say that the
rest frame of the cmb has zero peculiar velocity.
This indeed is what is usually assumed.

Inserting the Fourier expansion of $\psi$ and projecting out a
multipole leads, for an observer at the origin of coordinates,
 to
\be a_l^m=-2\pi i^l \sum \sk  \rfrac{aH}{k}^2
  j_l(kx\sub{em}) Y^m_l(\Omega\sk) \delta\sk \label{417} \ee
where $\Omega\sk$ is the solid angle in $\bfk$ space.
For a randomly placed observer the
coefficients $\delta\sk$ have random phases (except for the trivial
correlation implied by the reality condition $\delta\sk^*=\delta
_{-\mbox{\scriptsize\bf k}}$),
and this implies that the multipoles $a_l^m$ have a Gaussian distribution. The
variance of the distribution is (Peebles 1982b)
\be
\Sigma_l^2 = \pi \int_0^{\infty} \frac{{\rm d}k}{k} \, j_l^2
	\left( 2k/aH \right) \; \delta^2_H(k)
\ee
where $j_l$ is the spherical Bessel function. With
$\delta_H^2(k)\propto k^{n-1}$ this becomes
\be \Sigma_l^2=\frac{\pi}{2} \left[\frac{\sqrt\pi}{2}l(l+1)
\frac{\Gamma((3-n)/2)}{\Gamma((4-n)/2)}
\frac{\Gamma(l+(n-1)/2)}{\Gamma(l+(5-n)/2)}
\right]
\frac{\delta^2_H(H_0/2)}{l(l+1)} \label{422} \ee
The square bracket is equal to 1 for $n=1$. For $l\gg1$ and
$l\gg|n|$ it can be replaced by 1, if $\delta_H(k)$
is evaluated on the scale \eq{232} which dominates the
integral.

\subsection{The contribution of gravitational waves}

We have so far ignored gravitational waves. They correspond to a
metric perturbation $h_{ij}$ which
is traceless, $\delta^{ij} h_{ij}=0$, and transverse, $\pa_i h_{ij}=0$.
This means that each Fourier component is of the form
$h_{ij}=h_+ e^+_{ij} + h_\times e^\times_{ij}$, where
in a coordinate system where $\bfk$ points along
the $z$-axis
the nonzero
components of the polarisation
tensors are defined
by $e^+_{xx}=-e^+_{yy}=1$ and $e^\times_{xy}=e^\times_{yx}=1$.
The spectrum ${{\cal P}}_g$ of the
gravitational wave amplitude may be defined
by summing \eq{21} over all four components,
\be {{\cal P}}_g=2\rfrac{Lk}{2\pi}^3 \left( \langle|h_+(\bfk)|^2\rangle
+\langle|h_\times (\bfk)|^2\rangle \right)
\label{243}\ee

Each Fourier component satisfies the massless wave equation, which in comoving
coordinates is
\be \ddot h_{ij} +3H \dot h_{ij} +(k/a)^2 h_{ij} =0 \label{244}\ee
Well before horizon entry it has constant initial value.
For
scales entering the horizon after matter domination
its subsequent evolution is
\be h_{ij}(t)
=\left[3\sqrt\frac{\pi}{2} \frac{J_{3/2}(x)}{x\threehalf}
\right] h_{ij}(\mbox{initial}) \label{245}\ee
where $x=2k/(aH)$. Well after horizon entry
one has redshifting radiation.

As noted in \eq{duij}, the contribution of gravitational waves to the
velocity gradient is $\delta u_{ij}=\dot h_{ij}/2$.
By substituting this expression into \eq{delt} one can calculate the cmb
multipoles in terms of the initial amplitude,
 and hence calculate cmb spectrum $\Sigma_l^2$ in terms of
the initial gravitational wave spectrum
${{\cal P}}_g(k)$. Each gravitational wave gives its
dominant contribution as it enters the horizon, since its amplitude is
practically constant before that and redshifts away afterwards. As a result
the gravitational wave contribution to the cmb anisotropy cuts off below the
angular scale $\simeq 1^0$, corresponding to the smallest linear scale which
enters the horizon after decoupling $k\mone\sub{dec}=90h\mone\Mpc$. The
corresponding multipole cutoff is at $l\simeq k\mone\sub{dec} H_0/2\simeq 70$.

On scales well in excess of a degree ($l\lsim 10$),
the gravitational wave contribution
is given in the case of
a flat spectrum (${{\cal P}}_g(k)$ independent of $k$) by Starobinsky
\cite{star},
\be l(l+1)\Sigma_l^2=
\frac{\pi}{36}\left( 1+\frac{48\pi^2}{385}\right)
{{\cal P}}_g C_l \label{427} \ee
If one ignored the cutoff due to the redshift, the coefficient $C_l$ would
become $1$ in the limit $l\gg1$. Starobinsky gives the values $C_2=1.118$,
$C_3=0.878$  and $C_4=0.819$. Numerical calculation, including the effect of
the cutoff, shows that a value $C_l$ close to $1$ is indeed achieved for
$l\sim10$, before the cutoff takes effect (Figure 3).

For $l\gg1$ the above result is good also if ${{\cal P}}_g(k)$ has moderate
scale-dependence, provided that it is evaluated at the scale
\eq{232} which dominates the $l$th multipole.

Within a given inflationary model one can calculate ${{\cal P}}_g(k)$.
Defining the spectrum $n_g$ of the gravitational waves by ${{\cal P}}_g\propto
k^{n_g}$, one finds \cite{LL1}
\be n_g=-2\epsilon_1
\label{247}\ee
where $\epsilon_1$ is the same small
positive parameter appearing in \eq{221} for $n$.
Setting the coefficient $C_l$ in \eq{427} equal to 1, the
ratio of the
gravitational and density contributions is also given in terms of this
parameter, by \cite{LL1}
\be R\equiv \frac{\Sigma_l^2\mbox{(grav)}}
	{\Sigma_l^2(\mbox{density})}\simeq12\epsilon_1\label{430} \ee
In some models of inflation, $\epsilon_1$ is very
small, corresponding to a very small gravitational wave contribution with a
very flat spectrum. If $\epsilon_1$ is significant,
the potentially is typically of
the form $\phi^\alpha$ with $\alpha$ at least equal to 2 and often much
bigger (I include the behaviour $e^{A\phi}$ in this class since it
corresponds to the limit $\alpha\to \infty$). In that case one has
$-n_g=1-n>0$  and
\be
R\simeq 6 (1-n)
\label{249}\ee

Thus one's expectation from inflation is that {\it either}
gravitational waves are negligible {\it or} that their relative
magnitude is related to the spectral index by \eq{249}
(on scales in excess of a few degrees).
It should be noted that this expectation comes from particle physics,
being a consequence of the kind of potentials that
typically arise.
It is {\it not} generic to the idea of inflation {\it per se},
which in fact provides no relation whatever between $R$ and $n$
since the two parameters $\epsilon_1$ and $\eta_1$ can be chosen
independently.

\subsection{Observing the cmb anisotropy}

By analogy with \eq{corr}, one can define a temperature correlation
function,
\be C(\alpha)=\left\langle \frac{\Delta T(\bfe)}{T}
\frac{\Delta T(\bfe^\prime )}{T} \right\rangle \label{250}\ee
Here $\bfe$ and $\bfe^\prime $ specify the directions in which the
anisotropy is observed, and the
average goes over directions separated by
an angle $\alpha$. It is given in terms of the multipoles by
\be C(\alpha)=\sum_{l=2}^\infty Q_l^2 P_l(\cos\alpha) \label{251}\ee
where
\be
Q_l^2 \equiv \frac{1}{4\pi} \, \sum_{m=-l}^{+l} |a_l^m|^2
\label{ql2} \ee

Since one is not interested in very small scale fluctuations, it is
convenient to smear $\Delta T(\bfe)/T$ over a patch of sky with angular
size $\theta_f$, to obtain a `filtered' anisotropy
$\Delta T(\theta_f,\bfe )/T$. This cuts off the multipole expansion above
$l\sim\theta_f\mone$; to be precise, with
a Gaussian profile for the smearing
function, its effect is
\be a_l^m\to \exp \left[ -(2l+1)\theta_f/2\right] a_l^m \label{252}\ee
Associated with the smeared quantity is a correlation function
\be C(\theta_f,\alpha)=\sum_{l=2}^\infty
\exp \left[ -((2l+1)\theta_f/2)^2\right]  Q_l^2 P_l(\cos\alpha)
\label{corf} \label{253}\ee

A given
experimental setup typically measures something which can be directly related
to a suitably smeared $C$.
The simplest one is a single beam whose resolution can be represented
by a Gaussian profile. Averaged over the sky the anisotropy measured by such a
beam is
\be \sigma^2_T(\theta_f)\equiv
\left\langle \left[ \frac{\Delta T(\theta_f,\bfe)}{T} \right]^2
\right\rangle=C(\theta_f,0) \label{single} \ee
For more complicated setups involving two or three beam switching,
still with Gaussian profiles, the measured anisotropy is
\be \left\langle \left[ \frac{\Delta T(\theta_f,\bfe)}{T}
-\frac{\Delta T(\theta_f,\bfe^\prime )}{T} \right]^2 \right\rangle
= 2\left[ C(\theta_f,0)-C(\theta_f,\alpha) \right]
\label{double} \ee
and
\bea
&&\left\langle \left[ \frac{\Delta T(\theta_f,\bfe)}{T} -\frac12
	\frac{\Delta T(\theta_f,\bfe^\prime )}{T} -\frac12 \frac{\Delta
	T(\theta_f,\bfe^{\prime \prime })}{T} \right] ^2\right\rangle
\nonumber\\
&=&\frac32
	C(\theta_f,0)-2C(\theta_f,\alpha)+\frac12 C(\theta_f,2\alpha)
\label{triple} \eea
In the second expression,
$\bfe^\prime$ and $\bfe^{\prime \prime }$ lie on
opposite sides of $\bfe$, aligned on a great circle and each at an angular
distance $\alpha$. In a typical setup the `beam throw' $\alpha$ is of the same
order of magnitude as the `antenna resolution' $\theta_f$.

If we denote the left hand side of Eq.~(\ref{single}), (\ref{double}) or
(\ref{triple}) generically by $(\Delta T/T)^2$, the prediction in terms of
multipoles may be written
\be
 \rfrac{\Delta T}{T}^2 =\sum_{l=2}^\infty
F_l Q_l^2 \label{filter} \ee
where $F_l$ is a filter function.
For the single beam expression \eq{single} the filter keeps all multipoles
with $l\lsim\theta_f\mone$, but for the two- and three beam
expressions it keeps only multipoles with $l\sim \theta_f\mone$.

For a realistic experimental setup these Gaussian
expressions require modification.
Accurate filter functions for some currently mounted
observations are shown in Figure 3, which is reproduced from
\cite{cretal}.

\subsubsection*{Present observations}

Until recently the only published observation of the non-dipole
cmb anisotropy was that of the COBE group \cite{smet}.
 They give the quadrupole and
some information about higher multipoles, but because of cosmic variance
their most useful statistic is the rms anisotropy
on the scale $10^0$, the smallest that they explore.
\be
\sigma_T(\theta_f)=(1.10\pm.18)\times10^{-5}
\label{259}\ee
Assuming a Gaussian beam profile one can use this expression to
normalize the spectrum $\delta_H(k)$ of the density perturbation,
given the spectral index $n$ and the relative contribution
$R$ of gravitational waves. With $n=1$ and $R=0$ it gives
$\delta_H=(1.7\pm .3) \times 10^{-5}$.
Subsequent analysis \cite{newcobe}
using the correct beam profile and allowing for
incomplete sky coverage raises this to by about one standard deviation
to
\be
\delta_H=(2.0\pm .3)\times 10^{-5}
\label{260}
\ee

Since the ten degree measurement probes scales of order the Hubble
distance $k\mone\sim10^4\Mpc$,
 this provides a reasonable estimate of $\delta_H$ on that
scale even if $n\neq1$ is somewhat different from 1,
{\it provided} that gravitational waves are negligible.
Most inflationary models predict either that they will be negligible, or
that they will be given by \eq{249} (with $n<1$).
 In the latter case the normalisation \eq{260}
becomes
\be
\delta_H=(2.0\pm .3)\times 10^{-5} [1+6(1-n)]\mhalf
\label{260a}\ee

At the time of writing additional measurements on smaller scales
are being reported. They have bigger uncertainties than the COBE
measurement, and do not add substantially to what we learn by
combining this measurement with the results of galaxy surveys
in the manner discussed in the next section.

\section{The CDM and MDM Models of Structure Formation}

A model of large scale structure should simultaneously account for
at least half a dozen completely different types of data, exploring
a wide range of scales. Pure CDM with a flat spectrum ($n=1$) performs
rather impressively, but cannot actually fit all data within the
accepted uncertainties, and the same remains true if  one allows
$n$ to vary. Adding about 30\% of hot dark matter does work,
as far as theory and observation have been compared
\cite{mdm,LL1,bobqaisar}.

\subsection{The filtered density contrast}

At the present epoch the universe is highly inhomogeneous on small
scales. In order to use linear cosmological perturbation theory
one must therefore filter out the small
scales, by smearing each perturbation over a region of size $\gsim 100\Mpc$.
The same is true at epochs
in the relatively recent past, except that the comoving filtering
scale goes down. Only in the truly early universe is the universe
(presumably) homogeneous on all relevant scales.

The
filtering is done by means of a
`window function' $W(R_f,r)$, which is equal to 1 at $r=0$ and which falls off
rapidly beyond some radius $R_f$ \cite{kotu}.
Taking for definiteness the density contrast, the filtered quantity is
\be \delta(R_f,{\bfx})=\int W(R_f,|{\bfx}^\prime -{\bfx}|)
	\delta({\bfx}^\prime ) d^3 x^\prime  \label{261}\ee
and its spectrum is
\be {{\cal P}}_\delta(R_f,k)= \left[\widetilde W(R_f,k)/V_f\right]^2
	{{\cal P}}_\delta(k) \label{262}\ee
where
\be \widetilde W(R_f,k)=\int e^{-i \bf k\cdot x} W(R_f,r) d^3x
\label{263}\ee
and
\be V_f=\int W(R_f,r) d^3x \label{264}\ee
The filtered dispersion is
\be
\sigma^2(R_f)
= \int_0^{\infty} \left[\widetilde W(R_f,k)/V_f\right]^2
{{\cal P}}_\delta(k) \frac{{\rm d}k}{k}
\label{264a}
\ee

The quantity $V_f$ is the volume `enclosed' by the filter. It is convenient to
define the associated mass $M=\rho_0 V_f$, where $\rho_0$ is the
present mass density.
One normally uses $M$ instead of $R_f$ to specify the scale,
writing $\delta(M,\bfx)$ and $\sigma(M)$.

The two popular choices are the Gaussian filter
\bea
W(R_f,r) &=& \exp(-r^2/2 R_f^2) \label{gone} \\
V_f &=& (2\pi)\threehalf R_f^3 \\
\widetilde W(R_f,k)/V_f &=& \exp(-k^2 R_f^2/2) \\
M &=& 4.36\times 10^{12} h^2 (R_f/1\Mpc)^3 \msun
\label{gtwo} \eea
and the top hat filter which smears uniformly over a sphere of radius $R_f$
\bea W(R_f,r)&=&\theta(r-R_f) \\
V_f &=& 4\pi R_f^3/3\\
\widetilde W(R_f,k)/V_f &=& 3 \left( \frac{\sin(k R_f)}{(k R_f)^3}
-\frac{\cos(k R_f)}{(k R_f)^2} \right) \label{322} \\
M &=& 1.16\times 10^{12} h^2 (R_f/1\Mpc)^3 \msun
\label{273}\eea
The Gaussian filter is the most convenient for
theoretical calculations, but the top hat filter is widely used to as
a means of presenting data.

It is useful to write \eq{264a} in terms of the spectrum
$\delta_H^2$ of the primeval curvature. Using a Gaussian filter
it becomes
\be
\sigma^2(R_f)=\int^\infty_0 e^{-k^2 R_f^2}\left[
T^2(k) \delta_H^2(k) \rfrac{k}{aH}^4 \right] \frac{
\diff k}{k}
\ee
On the scales $k\mone\gsim 1\Mpc$ of most interest, the factor in square
brackets increases quite strongly with $k$, so that the entire integrand
peaks on the scale $k\mone\sim R_f$ and
\be
\sigma^2(R_f)\sim \calp_\delta(k)=
T^2(k)\rfrac{k}{aH}^4 \delta_H^2(k)
\label{fildis}\ee
with $k\mone=R_f$.
Thus, $\sigma(R_f)$ falls off as the smearing scale $R_f$ increases.

One can filter the other perturbations as well. The filtered peculiar velocity
is called the {\it bulk flow}. From \eq{159}, the dispersion of
a given component of the bulk flow is
\be
\sigma_v^2(R_f)=\int^\infty_0 e^{-k^2 R_f^2}\left[
T^2(k) \delta_H^2(k) \rfrac{k}{aH}^2 \right] \frac{
\diff k}{k}
\sim \rfrac{aH}{k}^2\calp_\delta
\ee
It too falls off with increasing scale, but not as fast as the
density contrast.

\subsubsection*{The bottom-up picture of structure formation}

In Figure 4 is shown the prediction for the linearly evolved
$\sigma(R_f)$, evaluated at the present epoch, for both
pure CDM and for MDM. From \eq{126a}, the pure CDM prediction
scales at earlier epochs like $(1+z)\mone$.

\begin{figure}
\vspace*{10cm}
\caption[Filtered density contrast]
{The dispersion $\sigma_0(M)$, of the linearly evolved
filtered density contrast evaluated at the present epoch,
is shown
for pure CDM and MDM, with COBE normalization. This figure is taken from
\cite{LL2}, which documents the precise input used.}
\end{figure}

The striking thing about the shape of $\sigma(R_f)$ is that it
decreases monotonically as $R_f$ increases. This
feature implies a `bottom-up' picture,
with structure forming on successively larger scales.
The epoch $z\sub{nl}(M)$,  when a significant fraction of
the mass in the universe collapses into objects with mass $M$,
is roughly the epoch when $\sigma(M)=1$. The linear evolution
$\sigma\propto a=(1+z)\mone$ given by \eq{126a}
then ceases to be valid, but if it {\it had}
remained valid the present value of $\sigma$ would have been
$(1+z\sub{nl})$. Thus
\be 1+z\sub{nl}(M)=\sigma_0(M) \label{274}\ee
where the $0$ denotes the present value of the {\em
linearly evolved } quantity.

Figure 4 leads therefore to the following general picture. First, a large
fraction of the matter in the universe collapses into objects
with the mass of a small galaxy, or even smaller. Then successively
large objects form, presumably gobbling up some or most of the smaller
objects. At the present epoch, collapse is only just beginning on the
scale $M\sim10^{15}\msun$ which corresponds to large clusters.

\subsection{Constraining the spectrum of the density
contrast}

\begin{table}
\centering
\begin{tabular}{|l|l|l|}
\hline \hline
OBSERVATION & NORMALISATION & REQUIRED $\sigma_8$ \\
\hline
\hline
COBE & $\sigma_T(10^0)=(1.1\pm.2)\times 10^{-5}$
& $1.15\pm.15$ \\
\hline
bulk flow & $\sigma_{19}=.37\pm.07$ & $.85\pm.15$ \\
\hline
cluster abundance & $\sigma_8=.46\pm.16$ & $.46\pm.16$ \\
\hline
galaxy correlation  & $E=1.21\pm.07$ & $E=.95$ \\
\hline
quasar abundance & $\sigma_1>.24$ at $z=4$ & $>.4$ \\
\hline \hline
\end{tabular}
\caption[Key data and their implied normalisation]
{Key data and their implied normalisation, taken from \cite{mymnras}.
The right hand column gives the value of $\sigma_8$ needed to fit each
observation with pure CDM and a flat spectrum, except in the
fourth row which gives $E$ defined by \eq{EXCESS} }
\end{table}

A full comparison of theory with observation requires the use of
numerical simulations, to follow the process of gravitational collapse
which occurs on each scale after cosmological perturbation theory breaks down.
It turns out, though, that the linear theory can be applied on a wide
variety of scales, so that one can obtain powerful constraints on the
parameters by considering  it alone.
Working from the top down, some of the most
important linear constraints are explained below.
\begin{itemize}
\item {\it The large scale cmb anisotropy} The COBE data
explores scales of order the size of the observable universe,
say $10^3 $ to $10^4\Mpc$.
\item {\it The bulk flow}
Smearing the galaxy peculiar velocities
over a sphere of radius tens of
Mpc to get what is called the {\it bulk flow},
one should (just) be in the linear regime.
In principle \cite{dekel}
one can observe the radial component of the bulk flow,
construct the corresponding potential  by integrating \eq{147b} radially,
reconstruct $\bfv$
and finally determine the density perturbation
$\delta\rho(\bfx)$ from \eq{147}.
In practice one has to use in addition
the number density contrast $\delta n/n$ of infrared galaxies
(for which the most complete data are available), assume that it
is equal to $b_I\delta\rho/\rho$ with a position independent bias factor
$b_I$, and estimate $\psi$ from \eq{148}.
\item {\it Galaxy cluster number density}
The average number density $n(>M)$ of clusters with mass bigger
than $M\sim10^{15}\msun$ gives information on a scale of order
$10h^{-1}$ Mpc. Within linear theory one can estimate $n(>M)$
by assuming that the matter in
regions of space where $\delta(M,\bfx)$ exceeds some critical
value $\delta_c$ of order 1 is bound into objects with mass
$>M$. The fraction of space occupied by such regions is
\be
f(>M)=\mbox{erfc}\rfrac{\delta_c}{\sqrt 2\sigma(M)}
\label{275}\ee
{}From this assumption Press and Schechter derived the formula
\cite{LL2}
\be
m\frac{\diff n(>M)}{\diff M} =\frac{\langle k^2\rangle}{12\pi^2
R_f} \nu e^{-\nu^2/2}
\label{276}\ee
where $\nu=\delta_c/\sigma(M)$ is the number of standard deviations that
$\delta_c$ represents, and
\be
\langle k^2(M)\rangle=\sigma\mtwo(M)\int^\infty_0
k^2e^{-k^2R_f^2}\calp_\delta(k)\frac{\diff k}{k}
\label{277}\ee
(This formula includes a more or less unmotivated  factor 2).
An alternative prescription is to identify each peak
in $\delta(M,\bfx)$ higher than $\delta_c$ with an object of mass
$>M$, which gives a roughly similar result.
Yet another method, which in principle is superior, is to run a
numerical simulation, which again gives roughly similar results.
\item {\it The shape of the galaxy correlation function}
The galaxy correlation function \eq{corr} can be used to probe the shape of
$\sigma(M)$ on scales between those explored by the last two items,
if the bias factor is taken to be scale independent. The result is
usefully quantified by giving the `excess power'
\be
\label{EXCESS}
E = 3.4 \frac{\sigma(25h^{-1} {\rm Mpc})}{\sigma(8h^{-1} {\rm Mpc}})
\ee
where the prefactor is chosen so that $E=1$ with pure
CDM and $n=1$. The observed value is somewhat higher.
\item {\it Quasar number density}
Given some astrophysics assumptions, the observed quasar abundance
can provide a lower limit on the density contrast at high redshift.
For instance, one estimate \cite{haehnelt} suggests that at $z=4$, the fraction
of mass bound into objects with $M>10^{13}\msun$ is at least
$1\times 10^{-7}$.
Using the Press-Schechter estimate
with $\delta_c=1.33$ \cite{LL2,LL1},
this fraction is
equal to 2 times \eq{275}, which gives $\sigma(M)>.24$ at $z=4$.
\end{itemize}
A subjective view \cite{mymnras} of the normalisation of $\sigma(R_f)$
on various scales by these observations
is given in the central column of Table 2,
with the notation $\sigma_r\equiv\sigma(r h\mone\Mpc)$. All values refer
to the present except the last one.

First let us see how pure CDM with a flat spectrum fares. The
value of $\sigma_8$ required to fits each observation is given in the
right hand column. The striking thing is that all of the observations
are consistent with this model to within a factor of 2. On the other
hand the observations are not actually fitted; normalising to
COBE the data fall below the prediction quite sharply as one moves down
through the tens of $\Mpc$ regime.

In Figure 5 is plotted the ratio of the MDM to the CDM prediction,
and one sees that it has a sharp drop of about the desired magnitude.
Also shown is the effect of tilting the pure CDM spectrum, by making
$n<1$. This too reduces the small scale power, but the drop
is not sharp enough to accomodate the galaxy correlation data
quantified by $E$.

\begin{figure}
\vspace*{10cm}
\caption[Comparison of CDM and MDM]
{The ratio of the MDM and pure CDM predictions shown in Figure 4
is shown, as a function of the filtering scale $r$. Also
shown is the result of replacing MDM by pure CDM with a tilted spectrum.
The figure is taken from \cite{LL2},
where the input is fully documented.}
\end{figure}

One can ask what constraints on
$n$ and the fraction $\Omega_\nu$ of HDM are allowed by the data,
if both are allowed to vary. Two investigations
have been performed \cite{mymnras,bobqaisar}. The second
analysis concludes that $.20\lsim\Omega_\nu\lsim .35$,
and that $n>.70$ (no gravitational waves) or
$n>.85$ (gravitational waves as in \eq{260a}), and the first
suggests even tighter constraints.
Similar bounds on $n$ also hold if one tries to rescue
pure CDM by throwing away the troublesome galaxy correlation data
\cite{LL1,LL2}
(justified perhaps on the ground that it assumes a scale independent
bias parameter).

These bounds on $n$ are extremely significant in the context of
inflation as we discuss in Section 6.

\subsection{Alternatives to the  MDM model}

We have seen that pure CDM roughly accounts for observation, but that
to get good agreement one needs a rather sharp reduction in power as one
goes down in scale from $\sim 30 h\mone \Mpc$ to $\sim 10 h\mone \Mpc$.
The MDM proposal is one way of giving such a reduction. What about
others?

Two have been tried so far.
One is to invoke a cosmological constant with $\Omega_\Lambda\simeq.6$,
leaving only $\Omega_m\sim.4$ in cold dark matter. While it goes in the
right direction, this fix does not seem to be as successful as MDM,
and is far less pleasing from a particle physics viewpoint.
The other is to leave the transfer function alone, and require that
the primeval power spectrum already has the sharp drop in power. Using
two inflationary fields one can put this (or practically any other feature)
into the spectrum, but such `designer' models are quite unmotivated from
a particle physics viewpoint.
Alternatives which have not been tried yet are to use some kind of
`warm' dark matter.

Instead of fixing the CDM model one can contemplate throwing it away.
Such a radical step seems unlikely to be needed, but
if one wishes to take it there are two possibilities. One is
to invoke topological defects  to be the seeds of
structure formation; the gauge cosmic strings discussed by Vachaspati
in his lectures might work if the dark matter is hot, and as with the
CDM model there is continuity as one goes up in scale from
galaxies to the scale $\sim H_0\mone$ observed by COBE.
The other is to use a primeval adiabatic or isocurvature density
spectrum, but to give it small scale power so that
the first objects to form have a mass of, say $1\msun$.
This would lead to a radically different scenario, with no {\it a priori}
connection between structure formation and the cmb anisotropy.

\section{Inflation}

It is widely supposed that the very early universe experienced an era of
inflation \cite{guth,kotu,CHAOTIC}.
 By `inflation' one means that the scale factor has positive
acceleration, $\ddot a>0$, corresponding to repulsive gravity.
During inflation $aH=\dot a$ is increasing, so that comoving scales are
leaving the horizon (Hubble distance) rather than entering it, and it is
supposed that at the beginning of inflation the observable universe was well
within the horizon.

Within the context of Einstein gravity,
inflation requires
negative pressure $p<\rho/3$ (\eq{52}).
 This is achieved if $p$ and $\rho$ are
dominated by a homogeneous,
slowly varying `inflaton field',
$\phi(t)$, because one knows from standard field theory that
\bea \rho&=& V+\frac12\dot\phi^2
\label{282}\\
p&=&-V +\frac12\dot\phi^2 \label{283}\eea
The condition for inflation is clearly $\dot\phi^2<V$.
An alternative is to modify Einstein gravity, with or without invoking
a scalar field, but in that case one can usually redefine the metric
so as to recover the Einstein gravity during inflation.
Until the end of this section I focus on the Einstein gravity case.

It is natural to assume that the field is sufficiently slowly varying
that the condition $p<\rho/3$ is amply satisfied, so that the potential
dominates $\rho$. One then has $\rho\simeq -p$, which from \eq{50}
makes $\rho$ practically constant,
\be
|H\mone\dot \rho/\rho|=1+p/\rho\ll1
\label{rhoslow}
\ee

\subsection{General features of inflation}

The most interesting prediction of inflation concerns the adiabatic
density perturbation and the gravitational waves, which at some level it
inevitably generates through vacuum fluctuations. Before examining them
let us look at the general picture.

\subsubsection*{Chaotic initial conditions?}

First let us ask how inflation is supposed to begin, and more generally
what happened before it. A simple proposal, and
the only one that
has been explored in detail, is Linde's proposal of `chaotic' initial
conditions \cite{CHAOTIC}. At some
initial epoch
a patch of the
universe which includes our own as a small part, is roughly homogeneous
and expanding roughly isotropically.
Presumably this initial epoch occurs at the Planck scale, with energy
density and expansion rate
$\rho\sim H\sim \mpl^4$, and it is reasonable to suppose that the size
of the patch is of order $H\mone$.

Inflation is supposed to be achieved in the following way.
The energy density comes from
the various scalar fields existing
in nature.
The initial values of these fields are random, subject to the constraint
that their potential $V(\phi,\psi,..)$ is of order $\mpl^4$
(it is supposed that their other contribution to $\rho$, coming from the
spacetime dependence of the fields, is of the same order rather than
much bigger.)
The
inflaton field $\phi$ is
distinguished from the non-inflaton fields $\psi,...$
by the fact that the potential is
relatively flat in its direction. Before $\phi$ has had time
to change much, the non-inflaton fields quickly adjust themselves
to minimise the potential at fixed $\phi$,
 after which inflation occurs as $\phi$ rolls slowly
down the potential.

An alternative might be to suppose that the universe is initially
{\it everywhere} roughly homogeneous and isotropic, with positive
curvature so that there is no boundary. The chaotic hypothesis is
more flexible in regard to model building, but otherwise the outcome is
much the same.

\subsubsection*{The $\Omega=1$ prediction}

We now come to the first virtue of inflation.
{}From \eq{7}, the time dependence of the density parameter is given by
\be
\Omega-1=\rfrac{K}{aH}^2
\label{284}\ee
Without inflation, the patch that we are considering
will typically either collapse in
a Hubble time or so ($\Omega\to \infty$) or will become practically empty
($\Omega\to 0$). With inflation, in contrast, $\Omega$ is driven
towards 1. Since $\Omega$ has the same value when observable universe
leaves the horizon as it does at present, this leads to the expectation
that {\it the present value of $\Omega$ is very close to 1}.
This conclusion might be avoided if the observable universe
leaves the horizon near the beginning of inflation,
which as we remark in a moment is difficult to arrange, or if the initial
value of $\Omega$ is rather finely tuned to zero \cite{lystom}.
Whether such escape routes are actually viable is not yet clear,
but rightly or wrongly the `prediction' $\Omega_0=1$
is widely regarded as one of the firmest
consequences of inflation.

To summarise, {\it we need inflation so that
the observable universe is sufficiently old, and is yet sufficiently
dense to form structure}. Naturally, the thought suggests itself that
non-inflating parts of the universe are common, but simply too
inhospitable for us to inhabit.

Let us note, for future reference, that
after $\Omega$ has been driven to a value fairly close to 1,
the expectation \eq{rhoslow}
that $\rho$ is slowly varying on the Hubble timescale
implies the same for $H$, leading to an approximately
exponential growth of
the scale factor, $a\propto e^{Ht}$.
We shall see that to
have a density perturbation of the right magnitude, $\rho$
at the end of inflation must be less than about
$(10^{16}\GeV)^4$, so if exponential inflation begins near
the Planck scale it must
last a very large number of Hubble times. This in turn would imply
that the epoch at which the observable universe leaves the horizon
occurs long after the beginning of inflation, because as we shall see
this epoch typically occurs
only 60 or so Hubble
times before the end of inflation.

\subsubsection*{Perfect classical homogeneity}

So far we need only be dealing only with a patch that is {\it roughly}
homogeneous and isotropic.
As in Section 1, the extreme homogeneity and isotropy of the
observable universe is `explained' by asserting that it is a very small
part of the whole, and that the whole has a non-fractal nature.\footnote
{The inflationary `explanation' of homogeneity and isotropy is
often stated incorrectly.
Namely, it is said that inflation `smooths out' or even `stretches' the
observable universe. This seems to me to be very misleading.
Before a scale leaves the
horizon during inflation the expansion of the universe is not important,
and we can think of field oscillations as particles. On
the reasonable assumption that interactions are unimportant
the
the occupation number of each momentum state is {\it time independent}.
After the scale leaves the horizon the
field inhomogeneity on that scale is frozen in because causal processes
cease to operate. This means that (except perhaps for rotational
velocity perturbations, which would decay anyway)
the observable universe is no more and no less homogeneous at the
beginning of inflation than it is now.}
 However, inflation makes this explanation particularly
attractive for two reasons. First,
the statement that the observable universe is a very
small part of the whole patch follows from the
exponential growth of $a$ combined with the constancy of $H$, provided
that the observable universe leaves the horizon at least a few Hubble
times after inflation begins \cite{LL2}.

Second and most crucially, we are free
to go all the way and assert that the observable universe is
{\it absolutely homogeneous} at the classical level, or in other words
that every Fourier mode of every field is in the vacuum state.
 In fact, not only
are we free to make this statement, it is mandatory on each scale
well before horizon exit. The reason is that
even one particle
per momentum state on such scales would spoil inflation,
by dominating the energy density \cite{LL2}.

If there is no inhomogeneity at the classical level, how is it
generated?  The answer, as we discuss in the next section,
 is that an inhomogeneity of well defined magnitude is
generated in a perfectly natural way by the quantum
fluctuation of the inflaton field.

\subsubsection*{Avoiding unwanted relics}

Without inflation, thermal equilibrium starts at $T\sim 10^{16}\GeV$.
Practically every particle is abundantly produced, and there can be many
phase transitions creating defects. Some of these relics are definitely
unwanted because they would be too abundant, notably examples being
magnetic monopoles produced at $T\gsim 10^{11}\GeV$ and
relic thermal gravitinos. Such relics can be avoided by inflation,
because the temperature after inflation
(reheat temperature) can be made low enough that the offending objects
are not produced. Of course, wanted relics might also be killed;
in particular structure forming cosmic strings need to have
energy per unit length of $\sim 2\times 10^{16}\GeV$, and
are ruled out by most inflationary models (but see \cite{latest}).

\subsubsection*{Summary}

There are two simple reasons for believing that the early universe
started with an era of inflation.
First, without inflation a given patch either collapses ($\Omega\to
\infty$) or becomes
practically empty ($\Omega\to0$) within a Hubble time, unless its density
parameter is very finely tuned to the value $\Omega=1$.
In contrast, inflation drives $\Omega$ towards 1 starting with an
arbitrary initial value. Second, inflation can postpone the epoch
of thermalisation of the universe, so that the temperature
is too low to produce cosmologically disasterous relics.

In addition, there is the more complicated issue of homogeneity and
isotropy. In order for inflation to occur, the energy density of the
inflating patch must not be dominated by small scale inhomogeneities.
In fact, on scales much less than the horizon (Hubble distance)
even one particle per momentum state
would be enough to spoil inflation; in other words the inflationary
hypothesis requires that the universe is perfectly homogeneous at the
classical level, on very small scales. As I now explain, the quantum
fluctuation on a given comoving scale then generates a well defined
inhomogeneity and anisotropy, which can be regarded as classical once
the scale leaves the horizon, and can explain the perturbation
the inhomogeneity and anisotropy of the observable universe.

\subsection{The spectrum of the density perturbation}

Inflation predicts a well defined adiabatic density perturbation in the
following way.
Well before horizon entry each Fourier mode of the inflaton field
is described by quantum free field theory in (practically) flat
spacetime. Its mean square quantum fluctuation can be calculated in a
straightforward way, being analogous to the Casimir effect for the
electromagnetic field which has been verified in the laboratory.
Corresponding to the fluctuation is a well defined adiabatic density
perturbation.
Primordial gravitational waves are also generated, through the quantum
fluctuation of the gravitational wave amplitude.
In this section I explain concisely how to do these calculations.

\subsubsection*{The slow roll conditions}

A homogeneous scalar field $\phi(\bfx,t)$
with minimal coupling to gravity has the
equation of motion
\be \ddot \phi+3 H\dot \phi +V^\prime (\phi) =0  \label{285}\ee

Practically all of the usually considered models of inflation satisfy three
conditions, usually termed the `slow roll' conditions.
Although the calculations can done without the slow roll conditions,
they become much more complicated.

The first slow roll condition is that
the motion of the field is overdamped, so that the `force'
$V^\prime $ balances the `friction term' $3H\dot\phi$,
\be
\dot{\phi} \simeq -\frac{1}{3H} V' \label{286}\ee
The second condition is that
\be
\epsilon \equiv \frac{m_{Pl}^2}{16\pi}
\left( \frac{V'}{V} \right)^2 \ll 1 \label{287}\ee
which means that the inflationary requirement $\dot\phi^2<V$ is well
satisfied and (assuming $\Omega=1$)
\be H^2\simeq \frac13 \frac{8\pi}{\mpl^2} V  \label{288}\ee
These two conditions imply that
$H$ is slowly varying, and that the scale factor increases more or less
exponentially,
\be a\propto e^{Ht} \label{289}\ee
The third condition is that
\be |\eta|\ll1  \label{290}\ee
where
\be \eta \equiv \frac{m_{Pl}^2}{8\pi} \frac{V''}{V} \label{291}\ee
It can be `derived' from the other two by differentiating
the approximation \eq{56} for $\dot\phi$ and noting that consistency with the
exact expression  \eq{52} requires $\ddot \phi\ll V^\prime $ is satisfied.
However
there is no logical necessity for the derivative of an approximation to be
itself a valid approximation, so this third condition is logically independent
of the others.

It should be noted that the first slow-roll condition is on a quite different
footing from the other two, being a statement about the {\em solution} of the
field equation as opposed to a statement about the potential that
defines this equation. What we are saying is that in the usually
considered models of inflation, one can show
that the first condition is an attractor solution, in a regime typically
characterised by the other two conditions, and that moreover reasonable initial
conditions on $\phi$ will ensure that this solution is achieved well before
the observable universe leaves the horizon.

\subsubsection*{Classical equations}

To define the perturbation $\delta \phi$ of the inflaton field, one has to
choose a slicing of space-time into spacelike hypersurfaces, just as was the
case for the perturbations $\delta\rho$, $\delta p$ and $\delta H$ that we
studied earlier. Since the inflaton field is supposed to dominate the
energy-momentum tensor, the momentum density vanishes if its spatial gradients
vanish \cite{bst}.  In other words, $\delta \phi$
vanishes if the hypersurfaces are chosen to be comoving!

In this situation one can proceed heuristically in the following way
\cite{lyth84,LL2}.
First one notes, from \eq{123}, that in the extreme slow roll limit
$p/\rho\to -1$ (corresponding to $\dot\rho\to 0$), the
curvature perturbation
$\calr$ of comoving hypersurfaces becomes infinite, if the comoving
density perturbation is finite. One should therefore define the
inflaton field perturbation $\delta\phi$ on
a family of hypersurfaces which remains undistorted in the
slow roll limit. The time displacement $\delta t$
of the comoving hypersurfaces
from the undistorted ones
is determined by the fact that $\delta\phi$ vanishes on the
latter,
\be
\delta t=-\delta\phi/\dot\phi
\ee
Finally, one can show either by brute force \cite{bardeen}
or by an elegant geometrical
argument \cite{LL2}
that the distortion of the comoving hypersurfaces caused by
this time displacement implies a curvature perturbation
$\calr=-H\delta t$, which in terms of the field perturbation is
\be
\calr=H\delta\phi/\dot \phi
\label{calr}\ee
Alternatively one can derive this equation without any
mention of the spatial geometry,
working entirely with the density perturbation
plus the definition \eqs{120}{90} of $\calr$ \cite{lyth84}.

Perturbing the inflaton field equation \eq{285} leads to
\be (\delta\phi\sk)\,\ddot{}+3H(\delta\phi\sk)\dot{}
+\left[\rfrac ka ^2 +V^{\prime \prime} \right] \delta
\phi\sk =0 \label{510} \ee
Until a few Hubble times after horizon exit \eq{58}
ensures that $V^{\prime \prime}$ can be dropped, so that the equation becomes
\be (\delta\phi\sk)\,\ddot{}+3H(\delta\phi\sk)\dot{}
+\rfrac ka ^2 \delta
\phi\sk =0 \label{511} \ee

A rigorous treatment \cite{mukhanov,sasaki}
is to define $\delta\phi$ on hypersurfaces
with zero curvature perturbation, \eq{calr} then being an
exact expression for the curvature perturbation of the comoving
hypersurfaces \cite{bardeen}.
 One can show that to first order in the cosmological
perturbations the field equation for
$\delta\phi$ is
\be
u^{\prime\prime}\sk+(k^2-z^{\prime\prime}u\sk)=0
\ee
where $u=a\delta\phi$, $z=a\dot \phi /H$ and a prime denotes
$a(\diff/\diff t)$.
Finally, one can show that this equation reduces to \eq{511}
if the slow roll conditions are satisfied. (The correction
can be calculated explicitly, and has been shown to be
small \cite{stly}.)

\subsubsection*{The quantum fluctuation}

Well before horizon entry $\delta\phi\sk $ is a massless
field living in
(practically) flat space-time, since its wavenumber $k/a$ is much
bigger than
the Hubble parameter $H$.
It can be quantised in the usual way so that its
quantum state  is labelled by the number of inflaton particles
 present with momentum $\bfk$, and we are assuming that there are no
particles, corresponding to the vacuum state.

Working in the
Heisenberg representation, $\phi\sk$ is associated with an operator
\be \hat\phi\sk(t)
=w_k(t) \hat a\sk+w^*_k(t)\hat a^\dagger_{-{\mbox{\scriptsize\bf k}}}
\label{518}  \ee
The annihilation operator $\hat a\sk$ satisfies the commutation
relation
\be [\hat a_{\mbox{\scriptsize \bf k }_1},
\hat a^\dagger_{\mbox{\scriptsize \bf k }_2} ]= \delta _{
\mbox{\scriptsize \bf k}_1, \mbox{\scriptsize \bf k}_2} \label{520} \ee
and the field satisfies the
commutation
relation,
\bea[\hat \phi(\bfx _1,t),\pdif{}{t}
\hat\phi(\bfx _2,t)]
&=&  i \delta^3(a \bfx_1-a \bfx _2)  \\
&=& ia^3\delta^3(\bfx_1-\bfx_2)
\eea
As a result the functions $w_k$ are given by
\be w_k(t)=a\mthreehalf(2k/a)\mhalf
e^{-i(\chi+(kt/a))} \label{519} \ee
The phase factor $\chi$ is arbitrary, and it has negligible
variation on the timescale $a/k$. The vacuum state is the one annihilated by
$\hat a\sk$, so the vacuum expectation value of the field perturbation is
\be \langle|\delta\phi\sk|^2\rangle=|w_k|^2
\label{521} \ee

To extend these results to the epoch of horizon exit and beyond, we have to
accept the validity of free field theory in curved space-time.
All we need to assume
is that there is a Heisenberg picture, in which operators satisfy the
classical equations of motion and state vectors are time independent. Then
$\hat\phi\sk$ continues to satisfy the field equation \eq{510}, and
\eqss{518}{520}{521} still hold, where $w_k$ is the solution of the field
equation reducing to \eq{519} well before horizon entry. As one easily checks,
the required solution is
\be w_k(t)=\frac{H}{(2k^3)\half }
\left(i+\frac{k}{aH}\right) e^{ik/aH} \label{523} \ee
A few Hubble times after horizon exit,
the vacuum expectation value is therefore
\be \langle|\delta\phi\sk|^2 \rangle=\frac{H^2}{2k^3}
\label{524} \ee

A measurement of the $\phi\sk$'s
will yield random phases, and a distribution of moduli whose dispersion is
given by \eq{524}. Accordingly the spectrum of the inflaton field, defined by
\eq{21}, is given a few Hubble times after horizon exit by
\be {{\cal P}}\half_\phi(k)=\frac{H}{2\pi} \label{529} \ee
Since $H$ is slowly varying on the Hubble
timescale it can be written,
a few Hubble times after horizon exit,
\be {{\cal P}}\half_\phi(k)=\frac{H_*}{2\pi} \label{526} \ee
where the star denotes the epoch of horizon exit, and
${{\cal P}}_\phi$ is evaluated a few Hubble times after horizon exit.

The spectrum of ${{\cal R}}$ is given by
\be {{\cal P}}_{{\cal R}}\half=\frac{H}{\dot\phi}{{\cal P}}_\phi\half
\ee
In Section 2
we learned that ${{\cal R}\sk}$ is constant after horizon
exit, so ${{\cal P}}_{{\cal R}}$ remains constant even though $H/\dot\phi$ and
${{\cal P}}_\phi$ might vary separately. It follows that as long as the scale
is
far outside the horizon,
\be {{\cal P}}_{{\cal R}}\half= \frac{H_*^2}{2\pi \dot\phi_*} \ee
where the star denotes the epoch of horizon exit.
Using \eq{56}, this
leads to \cite{lyth84}
\be \delta_H^2(k)= \frac{32}{75}\frac{V_*}{\mpl^4}\epsilon_*\mone
\label{512}\ee

\subsubsection*{Primordial gravitational waves?}

The gravitational wave amplitude
$h_{ij}$  is dynamically equivalent to a minimally coupled, massless
scalar field $\psi=(\mpl^2/16\pi)\half h_{ij}$.
Each non-zero Fourier component has the same
vacuum fluctuation as a massless field, with the above conversion factor. Thus
the spectrum of the gravitational waves is given by \eq{529},
\be {{\cal P}}_g(k)= 4\times \frac{16\pi}{\mpl^2} \rfrac{H_*}{2\pi}^2 \ee
Putting this expression into
\eq{427} and dividing it by \eq{422} gives the
ratio \eq{430} of the gravitational wave and density contributions to the cmb
anisotropy.

\subsubsection*{The classicality of the perturbation}

The quantum fluctuation of a field in flat spacetime cannot be regarded
as classical. The reason is that each Fourier mode is equivalent to a
harmonic oscillator in its ground state; the position and momentum of
such an oscillator cannot be accurately measured simultaneously
(unless the measured results are far out on the tail of the probability
distribution), and as a result the position cannot be well defined over
an extended period of time. Returning to the case of the field,
this means that a Fourier component cannot be defined with high accuracy
over an extended period of time.

After horizon exit though, a given Fourier component is no longer equivalent to
a harmonic oscillator and one can show that it {\it can} have a
 well
defined value \cite{lyth84,guthpi}. In this sense it is classical,
though of course one still has the `Schrodinger's cat' problem of how
and when it is supposed to have acquired this well defined value, drawn
from the quantum probablity distribution.

\subsubsection*{The inflationary energy scale}

Assuming that the adiabatic density perturbation dominates the
cmb anisotropy, we  saw earlier that gives $\delta_H\simeq 2.0\times 10^{-5}$
on the Hubble scale. \eq{512} therefore gives the energy
scale at the epoch when the observable universe leaves the horizon
during inflation,
\be
V\quarter_1= 7.3
\times 10^{16}\GeV \times
\left(\frac{\delta_H\epsilon_1}{2.0\times 10^{-5}}\right)\quarter
\label{vqua}
\ee
Since $\epsilon_1\ll 1$ we conclude that $V\quarter$ is less than a few
times $10^{16}\GeV$.
Since the energy density decreases with time, this is also an upper limit
on the energy density at the end of inflation, and on the reheat
temperature.

\subsection{Entering and leaving inflation}

In order to work out the observational consequences of
\eq{vqua} for particular inflationary models, we need
to do a little more work.

First, we need to ask how inflation ends. In the chaotic picture, two
possibilities exist concerning the minimum into which the non-inflaton
fields fall. The simplest possibility is that it corresponds to the true
vacuum; that is, that the non-inflaton fields have the same values as in the
present universe. Taking  the potential to be quadratic for simplicity,
it then has the form
\be
V=\frac12 m^2 \phi^2
\label{310}\ee
Inflation ends when the inflaton field starts to
execute decaying oscillations around its own vacuum value (zero in the
above example). This typically occurs when the slow roll conditions
fail,
\be \max\{\epsilon\sub{end},|\eta\sub{end}|\}\simeq 1 \label{510a} \ee
The hot Big
Bang ensues when the vacuum value has been achieved and the
decay products have thermalised. Models of this type have been widely
explored

The other possibility \cite{LIN2SC,LL2,LIN2SC2,latest,MML}
is that the minimum {\it at large fixed $\phi$}
corresponds to a
false vacuum, which dominates the $\phi$-dependent part of the potential.
In the quadratic case the potential is then
\be
V=V_0+\frac12 m^2\phi^2\simeq V_0
\label{311}\ee
where $V_0$ is the false vacuum energy density.
In this case inflation ends when the false vacuum is destabilized, as
$\phi$ falls through some critical value $\phi_c$.

Of course one can envisage more complicated possibilities. One is that
an early epoch of inflation gives way to a hot universe, which
finds itself in a false vacuum which is stabilized by the finite
temperature. The vacuum energy density $V_0$ dominates
when the temperature falls below $V_0\quarter$, and inflation occurs
for a more or less brief era as the field rolls slowly away from its
minimum, ending when it starts to oscillate around its true vacuum value.
 This way of beginning inflation is called `new inflation'
\cite{kotu}.
Apart from being more complicated, it is also even more
difficult to implement in
the context of sensible particle physics than the other two proposals.

\subsubsection*{Reheating}

In any model, the universe (or at least the constituents dominating the energy
density) must achieve thermal equilibrium at some point after inflation.
Because the first models of inflation invoked a thermal entry, this event is
usually called `reheating'.
The corresponding `reheat temperature' is the
biggest temperature ever achieved after inflation and plays a crucial role in
cosmology.
Unfortunately, no reliable estimate of it is known at present.
If inflation ends with a phase transition, reheating is
expected to be prompt.
If it ends with the inflaton field oscillating,
reheating may be long delayed because the couplings of the inflaton field to
other fields is then typically quite weak.

\subsubsection*{The epoch of horizon exit}

A first requirement for an inflationary model is that the observable
universe should be within the horizon at the beginning of inflation.
We therefore need to know the epoch when a given scale
leaves the horizon, something which is necessary also to calculate the
inflationary perturbations. Denoting it by a star, this epoch is given
by
\be a_* H_*=k \label{ashs} \ee
The epoch of horizon exit is therefore related to the
present magnitude of the scale in Hubble units by
\be
\frac{a_0H_0}{k} =\frac{a_0 H_0}{a_* H_*} \ee
Let us denote the end of inflation by a subscript `end' and the epoch of
reheating by `reh', assuming matter domination during the era
(if any) between these two epochs. Let us also assume that after
`reh' there is radiation domination, until the epoch `eq'
at which the dark matter density becomes equal to that of the radiation.
Throughout the history of the universe the
radiation energy density is proportional to $a\mfour$, that of
the matter is proportional $a\mthree$, and the total is proportional to
$H^2$. It follows that
\bea \frac{k}{a_0 H_0}&=&\frac{a_*}{a\sub{end}}
\frac{a\sub{end}}{a\sub{reh}}
\frac{a\sub{reh}}{a_0}
\frac{H_*}{H_0}\\
&=& e^{-N_*} \rfrac{\rho\sub{reh}}{\rho\sub{end}}\third
\rfrac{\rho_{0r}}{\rho\sub{reh}}\quarter
\rfrac{\rho_*}{\rho_0}\half
\eea
where $\rho_{0r}=(a\sub{eq}/a_0)\rho_0$ is the present radiation energy
density and $N_*$ is the number of Hubble times between horizon exit and the
end of inflation,
\bea N_*&\equiv&\ln(a_*/a\sub{end})\\
&\simeq& \int^{t\sub{end}}_{t_*} H dt \label{nsta}
\eea
It follows that
\be N_* = 62- \ln \frac k{a_0 H_0}
- \ln \frac{10^{16}\GeV}{V_*^{1/4}}
+ \ln \frac{V_*^{1/4}}{V\sub{end}^{1/4}} -\frac13 \ln\frac{V\sub{end}
\quarter}{\rho\quarter\sub{reh}}
\ee

This equation relates the three energy scales $V_*\quarter$,
$V\sub{end}\quarter$ and $\rho\sub{reh}\quarter$. The first two scales are
related by another expression for $N_*$, which follows from
\eqsss{nsta}{286}{287}{288}
\be
N_*=\frac{8\pi}{m_{Pl}^2} \int_{\phi\sub{end}}^{\phi_*} \frac{V}{V^\prime}
 d\phi =
\sqrt{\frac{4\pi}{m_{Pl}^2}} \; \left|\int_{\phi\sub{end}}^{\phi_*}
	\epsilon\mhalf d\phi \right| \label{nhub} \ee

The biggest scale that can be explored is roughly the present Hubble distance,
$a_0/k =H_0\mone=3000h\mone\Mpc$. I shall refer to the epoch of horizon exit
for
this scale as `the epoch when the observable universe leaves the horizon', and
denote it by the subscript 1. As displayed in \eq{vqua}, the COBE observations
require that $V_1\quarter\lsim 10^{16}\GeV$, with the equality holding in most
models of inflation. Also, \eq{nhub} gives in most models of inflation
$V_1\quarter\simeq V\sub{end}\quarter$. If reheating is prompt, we learn that
in most models the observable universe leaves the horizon about 62 e-folds
before the end of inflation. If reheating is long delayed $N_1$ could be
considerably reduced, being equal to 32 for the most extreme possibility of
$V\quarter\sub{reh}\sim 1000\GeV$ (corresponding to reheating just before the
electroweak transition which is presumably the last opportunity for
baryogenesis). For most purposes, however, one needs only the order of
magnitude of $N_1$.

The smallest scale on which the primeval perturbation can be probed at present
is around $1\Mpc$, and one sees that this scale leaves the horizon about 9
Hubble times after the observable universe. Discounting the case of very late
reheating, we conclude that in the usually considered inflationary models,
scales of cosmological interest leave the horizon 50 to 60 Hubble times before
the end of inflation.

\subsubsection*{The spectral index}

The spectral index $n$ of the density perturbation is
obtained by
differentiating \eq{512} with the aid of
\eqssss{286}{287}{288}{ashs}{289}
\be n=1+2\eta_1-6 \epsilon_1 \label{532a}
\ee
For definiteness I have taken to be the epoch
when the observable universe leaves the horizon.

\subsection{Specific models of inflation}

Of the many viable inflationary potentials that one could dream up, only a few
are reasonable from a particle physics viewpoint. Some of them are
described now.

\subsubsection*{True vacuum chaotic inflation}

If the potential is $V\propto\phi^{\alpha}$,
\eq{510a} shows that at the end of inflation
$\phi\sub{end}\simeq \alpha (\mpl^2/8\pi)\half$.
{}From \eq{nhub}
with $N_1=60$, $\phi_1^2\simeq120\alpha m_{Pl}^2/8\pi$. This leads to
$1-n=(2+\alpha)/120$  and $R=.05\alpha$, leading to the following
relation between the gravitational wave contribution and the spectral
index
\be R=6(1-n)-0.1 \label{ntor} \ee
For $\alpha=2$, 4, 6 and 10 one has
$1-n=.033$, $.05$, $.067$ and $.1$, and
$R=.10$, $.20$, $.30$ and $.50$.
Lazarides and
Shafi \cite{lazsha}
have shown that a potential of this kind follows from a class of
superstring-inspired gauge theories, with an
index $\alpha$ varying between $6$ and $10$.

\subsubsection*{False vacuum chaotic inflation}

The simplest model of false vacuum inflation is \cite{LIN2SC,LL2,LIN2SC2,
latest,MML}
\be V(\psi,\phi)=\frac14\lambda (\psi^2-M^2)^2+\frac12m^2\phi^2+
\frac12\lambda^{\prime}\phi^2\psi^2 \label{322b}\ee
The couplings
$\lambda$ and $\lambda^{\prime }$ are supposed to be somewhat less
than unity,
and in the numerical examples I set $\lambda=\lambda^\prime =0.1$.
For $\phi^2 >
\phi\sub{end}^2 = \lambda M^2/\lambda^\prime $, the potential for
the $\psi$ field has a local minimum at $\psi = 0$, in which the field is
assumed to sit (except for the quantum fluctuation). Inflation can occur as
$\phi$ slowly rolls down its potential
\be V(\phi)=\frac14 \lambda M^4 +\frac 12 m^2 \phi^2
\label{twsc} \ee
We will suppose that while observable scales leave the horizon the first term
dominates, since in the opposite case we recover the $\phi^2$
potential already considered.
When $\phi$ falls below $\phi\sub{end}$ this local minimum becomes a local
maximum, and a second order phase transition to the true vacuum occurs
which ends inflation and causes prompt reheating.

The requirement that the first term dominates
means that
\be \frac2\lambda \frac{m^2\phi^2_1}{M^4} \ll 1 \label{domi} \ee
Of the parameters $\epsilon$ and $\eta$ which are required to be small,
the second is independent of $\phi$,
\be \eta=\frac4\lambda X^2 \ee
where
\be X^2\equiv \frac{m_{Pl}^2}{8\pi} \frac{m^2}{M^4} \ee
The ratio $X$ must therefore be significantly less than 1. From
\eq{nhub},
\be \phi_1=\sqrt{\lambda/\lambda^\prime } \, M e^{N_1\eta} \ee
Consistency with \eq{domi} requires roughly $\eta\lsim .1$.
The other small quantity $\epsilon_1$ is given by
\be \epsilon_1=\frac12 \frac\lambda{\lambda^\prime }
\frac{8\pi}{m_{Pl}^2}M^2 \eta^2 e^{2N_1\eta} \ee
and \eq{domi} requires $\epsilon_1\ll \eta$.
It therefore follows that the spectral index $n$ is {\em bigger} than
$1$ in the two-scale model.

Setting $\lambda=\lambda^\prime =.1$ and imposing the COBE normalisation
$\delta_H=1.7 \times 10^{-5}$
determines all of the parameters in terms of
$m$. For $m=100\GeV$ one has $M=4\times10^{11}\GeV$ leading to
$\eta=10\mfour$ and $\epsilon_1
=10^{-23}$. The gravitational waves are
absolutely negligible, and
the spectral index is extremely close to 1.
The maximum value of $m$ permitted by \eq{domi} is roughly
$m=10^{13}\GeV$, giving
$M=2\times 10^{16}\GeV$, $\eta=.07$ and $\epsilon_1
=10\mthree$. The gravitational waves are still negligible, but
$n=1.14$, significantly {\em bigger} than 1.

So far we have considered a single scalar field $\psi$, possessing the discrete
symmetry $\psi\to-\psi$. When inflation ends, domain walls will form along the
surfaces in space where $\psi$ is exactly zero, so to make the model
cosmologically viable one would have to get rid of the walls by slightly
breaking the symmetry. However, it is clear that one can easily consider more
fields, with continuous global or gauge symmetries. In particular, if $\psi$
is complex one can use the same potential with the replacement
$\psi^2\to|\psi|^2$. This leads to exactly the same inflationary model, but
now global strings are formed instead of domain walls.

The case $m\sim 100\GeV$ and $M\sim10^{11}\GeV$ is particularly
interesting because these are rather natural scales in particle physics.
One possibility is to identify $\psi$ with the Peccei-Quinn field,
and it might also be possible to realise the model in the context of
superstrings \cite{latest}

\subsubsection*{Natural Inflation.}

The `natural inflation' model \cite{freese,adams} has the potential
\be
V(\phi) = \Lambda^4 \left( 1 \pm \cos (\phi/f) \right)
\label{562} \ee
where $\Lambda$ and $f$ are mass scales.
In this model
\bea \epsilon&=& r \mone\tan^2\frac{\phi}{2f} \\
\eta&=& -r \mone\left( 1-\tan^2\frac{\phi}{2f} \right)
\eea
where $r\equiv (16\pi f^2)/\mpl^2$. Since $\epsilon-\eta=r\mone$ one must have
$r\gg 1$ to be in the slow-roll regime anywhere on this potential.
This requires that $f$ is at the Planck scale, suggesting a connection
with superstrings which has been explored in \cite{adams}.

In this model inflation has to begin with $\phi$ close to zero.
If there is a preceding thermal era, one might expect
$\phi$ to be chosen randomly so that a value close to zero is certainly
allowed. With or without a thermal era, additional fields have to be
involved to have a chaotic beginning at the Planck scale.

Inflation ends with an oscillatory exit at the epoch given by \eq{510a},
$\tan(\phi\sub{end} /2f)\simeq r\half$ which is of order 1. Using this result,
\eq{nhub} with $N_1=60$ gives $\phi_1/2f\simeq\exp({-60/r})$, leading to
\bea \epsilon_1&=& \frac 1{r} e^{-120/r} \\
\eta_1&\simeq& -1/r\\
1-n&\simeq& 2/r
\eea
Thus, natural inflation makes the gravitational waves negligible but tends to
give a significantly tilted spectrum.  The observational bound
$n>.7$ implies that $r>6$.

\subsubsection*{$R^2$ inflation}

The Lagrangian density for Einstein gravity is just $-(\mpl^2/16\pi)
R$, where $R$ is the spacetime curvature scalar.
Perhaps the simplest model of inflation is to replace $R$ by
$R+R^2/(6M^2)$. By redefining the metric one can then recover Einstein
gravity, at the expense of modifying the rest of the Lagrangian
{\it and} adding a completely new field $\phi$ with a potential
\be V(\phi)=\frac{3\mpl^2 M^2}{32\pi}
\left[ 1-\exp\left(-\rfrac{16\pi}{3\mpl^2}\half \phi\right) \right]^2
\ee
 In the regime $\phi\gsim m_{Pl}$, the
potential $V(\phi)$ satisfies the slow-roll conditions so that inflation
occurs. The non-gravitational sector is irrelevant during inflation, and if
the $R^2$ term quickly becomes negligible afterwards we have a model in which
inflation is driven by a scalar field, without any modification of gravity.
What has been achieved is to motivate the otherwise bizarre form of the
inflationary potential.

In the regime $\phi\gsim\mpl$ where they are small, the parameters appearing
in the slow-roll conditions are
\bea
\eta&=& - \frac43
\exp\left( -\sqrt\frac23 \frac{\sqrt{8\pi}}{\mpl}\phi\right)\\
\epsilon&=&\frac34 \eta^2
\eea
There is an oscillatory exit to inflation at the epoch given by \eq{510a},
$\phi\sub{end}\sim (\mpl^2/8\pi)\half $. From \eq{nhub} with $N_1=60$, \be
\phi_1\simeq 5 \frac{\mpl}{\sqrt{8\pi}} \ee leading to $\eta_1\simeq -.02$ and
$\epsilon_1\sim 10\mfour$. Thus the spectral index is $n=0.96$,
but the gravitational wave contribution is completely
negligible.

Justification for this model of inflation has recently been claimed in
the context of superstrings \cite{ovrut}

\subsubsection*{Extended inflation}

`Extended inflation' models are based on modifications
of Einstein gravity of the Brans-Dicke type
\cite{LaSteinhardt}. After recovering Einstein
gravity by redefining the metric, one typically
has an inflationary  scalar field, with the potential
\be
V=V_0\exp\left(\sqrt{\frac{16\pi}{p\mpl^2}} \phi\right)
\ee
Irrespective of when inflation ends, one finds
$-n_g=1-n(=2/p)$, and the `canonical' gravitational wave ratio
$R\simeq6(1-n)$ that we discussed earlier.

Inflation in these models ends with a first order phase transition
(caused, in the
Einstein gravity picture, by the modification of the non-gravitational
Lagrangian induced by the metric transformation).
The bubbles that it generates must
not cause an unacceptable
microwave background anisotropy, which generally leads to the constraint
$n \lsim 0.75$ \cite{LW1,LL1}, in conflict with the observational
bound $n>.85$ that we noted in Section 5.
One needs to make these models quite complex in order to avoid this
difficulty \cite{Laycock,LIN2SC2}.

\subsection{Summary and prospects}

Particle physics, especially in the context of superstrings, is
beginning to hint at several viable models of inflation. Each of them
makes a distinctive prediction regarding the tilt of the
density perturbation spectrum, which will allow one to discriminate
between them in the near future provided that some variant of the CDM
model continues to fit the observations. One will then, for the first
time, have an observational window on energy scales not far below the
Planck scale.

\end{document}